\documentclass[11pt,a4paper]{article}
\pdfoutput=1
\usepackage{jcappub}
\usepackage{rotating}
\usepackage{array}
\usepackage{amsmath}
\usepackage[normalem]{ulem}
\usepackage{slashed}
\usepackage{booktabs}
\usepackage[pdftex,table]{xcolor}
\usepackage{units}

\usepackage{multirow}

\newcommand{\erf}{\text{erf}}

\arxivnumber{DESY-17-113, HEPHY-PUB 991/17}

\title{Exploring light mediators with low-threshold direct detection experiments}

\author[a,b]{Felix Kahlhoefer,}
\author[c]{Suchita Kulkarni}
\author[a]{and Sebastian Wild}

\affiliation[a]{Deutsches Elektronen-Synchrotron (DESY), Notkestrasse 85, D-22603 Hamburg, Germany}
\affiliation[b]{Institute for Theoretical Particle Physics and Cosmology (TTK), RWTH Aachen University, \\ D-52056 Aachen, Germany}
\affiliation[c]{Institut f\"ur Hochenergiephysik,  
\"Osterreichische Akademie der Wissenschaften, \\ Nikolsdorfer Gasse 18, 1050 Wien, Austria}

\emailAdd{felix.kahlhoefer@desy.de}
\emailAdd{suchita.kulkarni@oeaw.ac.at}
\emailAdd{sebastian.wild@desy.de}

\abstract{
We explore the potential of future cryogenic direct detection experiments to determine the properties of the mediator that communicates the interactions between dark matter and nuclei. Due to their low thresholds and large exposures, experiments like CRESST-III, SuperCDMS SNOLAB and EDELWEISS-III will have excellent capability to reconstruct mediator masses in the MeV range for a large class of models. Combining the information from several experiments further improves the parameter reconstruction, even when taking into account additional nuisance parameters related to background uncertainties and the dark matter velocity distribution. These observations may offer the intriguing possibility of studying dark matter self-interactions with direct detection experiments.}

\keywords{dark matter theory, dark matter experiments, cosmology of theories beyond the SM, particle physics - cosmology connection}

\begin{document}
\maketitle
\flushbottom

\section{Introduction}

Dark matter (DM) direct detection experiments are conventionally interpreted in terms of effective operators, which parametrize the details of the underlying interactions of the DM particle in terms of an effective suppression scale $\Lambda$~\cite{Fan:2010gt,Fitzpatrick:2012ix,Anand:2013yka,Dent:2015zpa}. This is a good approximation as long as the particle mediating the interaction is heavy compared to the typical momentum transfer in direct detection experiments, which is of order $1\text{--}100\,\mathrm{MeV}$. Nevertheless, it is perfectly conceivable that the mediator mass is close to or below this scale, so that the effective operator description is no longer valid~\cite{Foot:2004pa,Fornengo:2011sz,Chu:2011be,Hooper:2012cw,Kaplinghat:2013yxa,Li:2014vza}. This scenario has for example been considered in the context of self-interacting DM~\cite{deLaix:1995vi,Spergel:1999mh}. In these models the exchange of light mediators induces large DM self-scattering rates~\cite{Ackerman:mha,Feng:2009mn,Buckley:2009in,Feng:2009hw,Loeb:2010gj,Aarssen:2012fx,Tulin:2012wi,Tulin:2013teo,Kaplinghat:2015aga,Bringmann:2016din,Kahlhoefer:2017umn}, which can potentially resolve the tension between the predictions of collisionless cold DM and observations on small astrophysical scales~\cite{Buckley:2009in,Feng:2009hw,Feng:2009mn,Loeb:2010gj,Zavala:2012us,Vogelsberger:2012ku,Buckley:2014hja}.

Current experimental limits e.g.\ from searches for rare decays~\cite{Hewett:2012ns,Dolan:2014ska,Krnjaic:2015mbs} put strong constraints on the mediator coupling to Standard Model (SM) particles. Nevertheless, the smallness of the mediator mass leads to a huge enhancement of direct detection cross sections, so that an observation of DM scattering may be possible in spite of the small couplings. In fact, if the mediator has sizeable couplings to DM, direct detection experiments can probe regions of parameter space inaccessible to other low-energy searches. As the recoil spectrum depends sensitively on both the DM and mediator mass, given an observation of DM scattering via a light mediator with sufficient statistics at a direct detection experiment, it may be possible to reconstruct both masses.

Cryogenic direct detection experiments, such as SuperCDMS~\cite{Agnese:2015nto,CDMSlitedata,Agnese:2016cpb}, CRESST~\cite{Angloher:2015ewa,Angloher:2015eza,Angloher:2017zkf,Strauss:2016sxp} and EDELWEISS~\cite{Armengaud:2016cvl,Hehn:2016nll,Arnaud:2017usi}, are particularly well-suited for this task due to their excellent energy resolution and low threshold~\cite{Peter:2013aha,Cooley:2014aya}. In fact, a low energy threshold is even more important in the case of a light mediator than for a heavy one, because the recoil spectrum falls even more steeply and therefore the sensitivity can be considerably improved by lowering the threshold~\cite{An:2014twa}. The excellent energy resolution, on the other hand, makes it possible to extract the maximum amount of information on the particle physics properties of DM from a successful discovery. In other words, cryogenic detectors are not only well-suited to explore models with light DM particles (see e.g.~\cite{Gelmini:2016emn}), but also to probe light \emph{mediators}.

The projected progress for the low-threshold technology implies that parameter points that are currently consistent with all experimental constraints may predict up to thousands of events in near-future detectors. In this paper we study the amount of information that can be extracted from such a signal, taking into account background uncertainties, astrophysical uncertainties and degeneracies with other particle physics parameters. We demonstrate that cryogenic experiments can probe the mediator mass precisely in the regions of parameter space relevant for DM self-interactions, potentially enabling us to infer the behaviour of DM on astrophysical scales with laboratory experiments.

Direct detection experiments in the context of self-interacting DM have been studied previously~\cite{Vogelsberger:2012sa,Kaplinghat:2013yxa,Chen:2015bwa,Kahlhoefer:2017umn}, most notably in ref.~\cite{DelNobile:2015uua}. Our work differs from these earlier studies in that we do not attempt to derive existing constraints but rather to explore the potential of future low-threshold detectors to infer the properties of self-interacting DM. For this purpose, we implement several present and future direct detection experiments in a realistic and efficient manner, in order to perform parameter reconstruction with a number of nuisance parameters. For similar studies in the context of effective operators see refs.~\cite{Strigari:2009zb,Pato:2011de,Kavanagh:2013wba,Feldstein:2014gza,Gluscevic:2015sqa,Kahlhoefer:2016eds,Rogers:2016jrx,Roszkowski:2016bhs,Roszkowski:2017dou}.

This paper is structured as follows. In section~\ref{sec:mediator} we discuss the phenomenology of direct detection experiments in the presence of light mediators. We review current and proposed low-threshold experiments and calculate their sensitivity to long-range interactions in comparison to conventional direct detection experiments. Section~\ref{sec:reconstruction} focusses on the potential of low-threshold experiments to determine the particle physics parameters of the DM particle and its interactions. We discuss the impact of experimental, theoretical and astrophysical uncertainties, introduce suitable nuisance parameters to represent them and assess their impact on our results. Finally, in section~\ref{sec:SIDM} we connect our results to the idea of self-interacting DM. Additional details are provided in appendices~\ref{app:experiments} and~\ref{app:astro}.

\section{Direct detection with light mediators}
\label{sec:mediator}

We consider a DM particle of mass $m_\text{DM}$ scattering off nuclei via the exchange of a mediator with mass $m_\text{med}$. Throughout this paper we will focus on the case that the mediator has spin-independent couplings to both nucleons and DM. The differential event rate with respect to recoil energy $E_\mathrm{R}$ for DM scattering off a given target isotope $T$ with mass $m_T$ and mass fraction $\xi_T$ is then given by
\begin{align}
\frac{\text{d}R_T}{\text{d}E_\mathrm{R}} = \frac{\rho_0 \, \xi_T}{2 \pi \, m_\text{DM}} \frac{g^2 \, F_T^2(E_\mathrm{R})}{\left( 2 \, m_T \, E_\mathrm{R} + m_\text{med}^2 \right)^2} \, \eta ( v_\text{min} (E_\mathrm{R}))
\label{eq:dRdE}
\end{align}
with $\rho_0 = 0.3\,\text{GeV}/\text{cm}^3$ being the local DM density.

As long as the assumption of spin-independent interactions holds, the functional form of the differential event rate does not depend on the spin of the DM particle or the mediator nor on whether or not the DM particle is its own anti-particle. The numerical pre-factors, however, may differ for these different scenarios. We assume that these pre-factors have been absorbed into the definition of the effective low-energy coupling $g$, i.e.\ we take $g$ to be defined via eq.~(\ref{eq:dRdE}). The precise definition of $g$ in terms of the fundamental parameters of the specific models that we discuss will be provided below.

In eq.~(\ref{eq:dRdE}) the factor $F_T^2(E_\mathrm{R})$ denotes the nuclear response function, which depends on the ratio of the mediator couplings to neutrons and protons, $f_n / f_p$. We parametrise this ratio via $\theta \equiv \arctan f_n / f_p$, such that models with $f_n = 0$ (such as a dark photon with kinetic mixing~\cite{Foot:2004pa,Feng:2009mn,An:2014twa}) have $\theta = 0$, whereas models with $f_n = f_p$ (such as a light scalar mixing with the Higgs boson~\cite{Kaplinghat:2013yxa,Kouvaris:2014uoa,Bernal:2015ova,Krnjaic:2015mbs}) have $\theta = \pi / 4$. In the limit of zero momentum transfer the nuclear response function is then given by $F_T^2(0) = (Z_T \cos \theta + (A_T - Z_T) \sin \theta)^2$, where $Z_T$ and $A_T$ denote the charge and mass number of the target isotope $T$. For non-zero momentum transfer, $F_T^2(E_\mathrm{R})$ decreases due to a loss of coherence, which we parametrise by the standard Helm form factors~\cite{Lewin:1995rx}.

The final factor in eq.~(\ref{eq:dRdE}) denotes the velocity integral, which is given by
\begin{equation}
 \eta(v_\text{min}) = \int_{v_\text{min}}^\infty \mathrm{d}^3v f(\mathbf{v})/v \; ,
\end{equation}
where $v_\text{min} = \sqrt{\frac{m_T E_\text{R}}{2 \, \mu_{T}^2}}$. Unless stated otherwise, we assume the velocity distribution $f(\mathbf{v})$ to be given by an isotropic Maxwell-Boltzmann distribution in the Galactic rest frame with $v_0 = 220 \text{ km/s}$, cut off at the Galactic escape velocity $v_\text{esc} = 544 \text{ km/s}$ and transformed into the solar rest frame with $v_\text{obs} =232 \text{ km/s}$.

For the purpose of this work, we will be most interested in the impact of the term $\left( 2 \, m_T \, E_\mathrm{R} + m_\text{med}^2 \right)^2 = (q^2 + m_\text{med}^2)^2$ in the denominator, where $q$ denotes the momentum transfer in the scattering process. This momentum transfer is bounded by \mbox{$q^\text{min} < q < q^\text{max}$}, where \mbox{$q^\text{min} \equiv \sqrt{2 m_T E_\text{th}}$} in terms of the low-energy threshold $E_\text{th}$ and \mbox{$q^\text{max} \equiv 2 \, \mu_T (v_\text{esc} + v_\text{obs})$} with \mbox{$\mu_T \equiv m_\text{DM} \, m_T / (m_\text{DM} + m_T)$}. If $q^\text{max}$ is small compared to $m_\text{med}$, we recover the limit of contact interactions conventionally considered in the effective operator approach. Conversely, if $m_\text{med}$ is negligible compared to $q^\text{min}$, scattering with large momentum transfer is suppressed by an additional factor of $1/q^4$, leading to very steeply falling recoil spectra. In the intermediate regime we may hope to infer the properties of the mediator from the detailed shape of the recoil spectra. These interesting mediator masses are typically in the MeV range. For example, for $m_\text{DM} = 4\,\text{GeV}$ a germanium experiment ($m_T \approx 68\,\text{GeV}$) with $E_\text{th} = 100\,\text{eV}$ has $q^\text{min} \approx 3.7\,\text{MeV}$ and $q^\text{max} \approx 20.7\,\text{MeV}$.

In real direct detection experiments, the detected recoil energy $E_\mathrm{D}$ may differ from the true physical recoil energy $E_\mathrm{R}$ due to the finite energy resolution of the detector. Assuming this resolution to be described by a Gaussian distribution with energy-dependent standard deviation $\sigma(E_\mathrm{R})$, we can calculate the probability for a scattering event with energy $E_\mathrm{R}$ to be observed in the interval $E_1 \leq E_\mathrm{D} \leq E_2$:
\begin{equation}
 p(E_\mathrm{R}, E_1, E_2) = \frac{1}{2}\left[\erf\left(\frac{E_2 - E_\mathrm{R}}{\sqrt{2} \sigma(E_\mathrm{R})}\right) - \erf\left(\frac{E_1 - E_\mathrm{R}}{\sqrt{2} \sigma(E_\mathrm{R})}\right)\right] \; ,
\end{equation}
where $\erf$ denotes the error function. Given the total exposure of an experiment $\kappa(E_\mathrm{R})$, which may again depend on the recoil energy, we can then calculate the total number of events expected in the interval $[E_1, E_2]$:
\begin{equation}
 N(E_1, E_2) = \sum_T \int p_T(E_\mathrm{R}, E_1, E_2) \, \kappa_T(E_\mathrm{R}) \, \frac{\text{d}R_T}{\text{d}E_\mathrm{R}} \mathrm{d}E_\mathrm{R} \; ,
\end{equation}
where the sum is over all isotopes $T$ in the target, weighted with appropriate factors $\xi_T$, and we allow both $p$ and $\kappa$ to depend on the isotope.

The two most sensitive cryogenic direct detection experiments are CRESST-II~\cite{Angloher:2015ewa,Angloher:2017zkf} and CDMSlite~\cite{Agnese:2015nto}. In the near future CRESST-III~\cite{Angloher:2015eza,Strauss:2016sxp} and SuperCDMS SNOLAB~\cite{Agnese:2016cpb} plan to significantly improve sensitivity.\footnote{A proposal for a similar effort with optimized EDELWEISS-III detectors was very recently put forward in ref.~\cite{Arnaud:2017usi}. The suggested approach is very similar to the one employed by the SuperCDMS collaboration and we expect qualitatively similar results.} We describe our implementation of these experiments in appendix~\ref{app:experiments}. For comparison, we also consider bounds from Xenon1T~\cite{Aprile:2017iyp}, as well as projections for the future sensitivity of LZ~\cite{Akerib:2015cja}. We note that all projections have similar time scales and correspond to the sensitivity that may be achievable within the next five to ten years.

\begin{figure}
\centering
\includegraphics[width=0.6\textwidth]{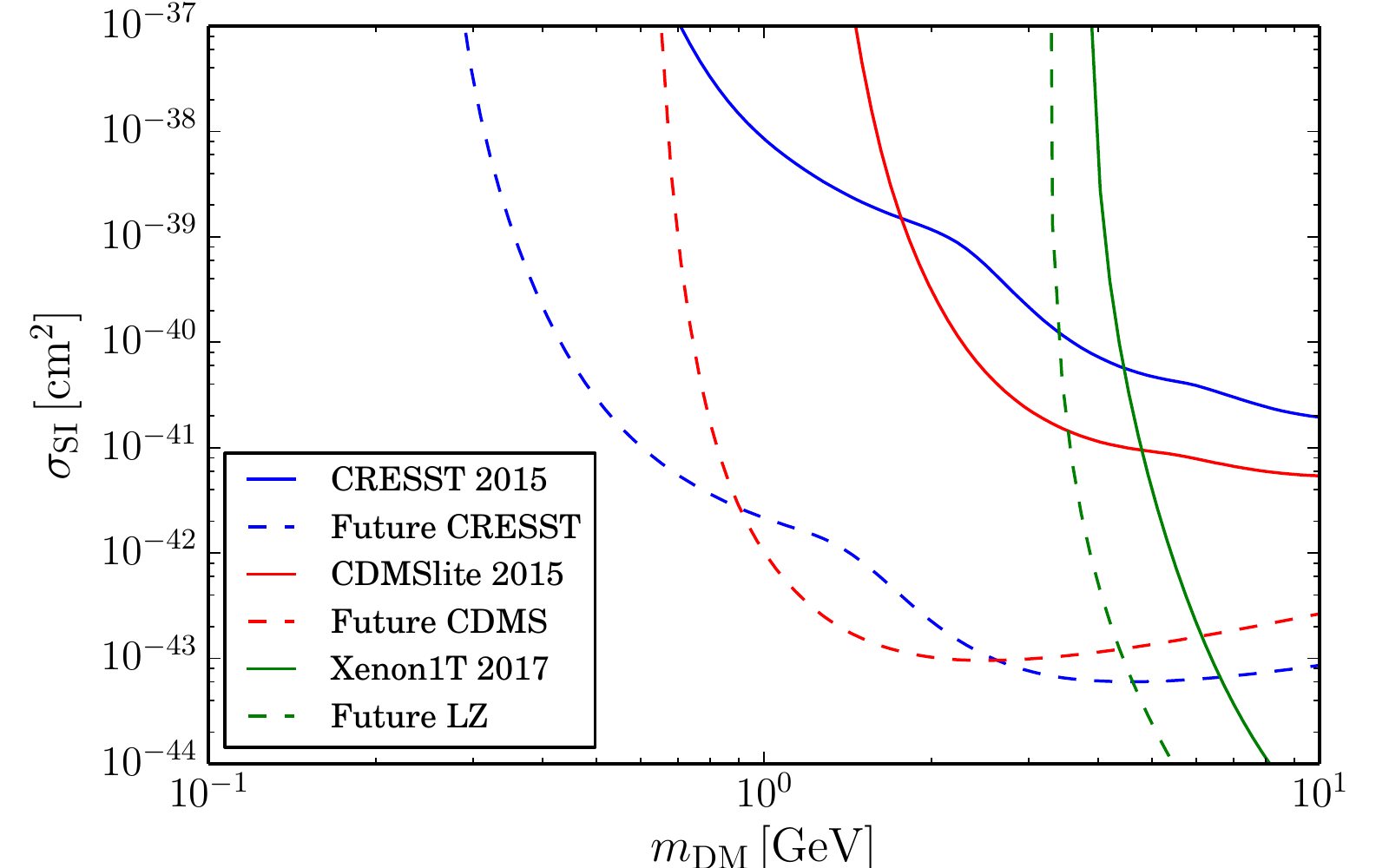}
\caption{\label{fig:SI} Exclusion limits (at 90\% CL) and projected sensitivities of various present and future direct detection experiments for standard spin-independent scattering.}
\end{figure}

As a validation of our implementation, we show in figure~\ref{fig:SI} the (projected) constraints on standard spin-independent interactions that we obtain from present and future direct detection experiments. For this purpose, we take $m_\text{med} \gg q^\text{max}$, set $\theta = \pi / 4$ and $g^2 = 2 \pi \, \sigma_p \, m_\text{med}^4 / \mu_p^2$, where $\mu_p \equiv m_\text{DM} \, m_p / (m_\text{DM} + m_p)$ and $m_p$ denotes the proton mass. Our recalculation of existing constraints is in good agreement with the respective published exclusion limits. For SuperCDMS and CRESST-III we have chosen a rather conservative low-energy threshold of $100\,\mathrm{eV}$ in both cases, so that our projected exclusion limits are somewhat weaker than the official projections for small DM masses. Nevertheless, we find good agreement at larger DM masses, where the precise value of the threshold does not matter.

\begin{figure}
\centering
\includegraphics[width=0.48\textwidth]{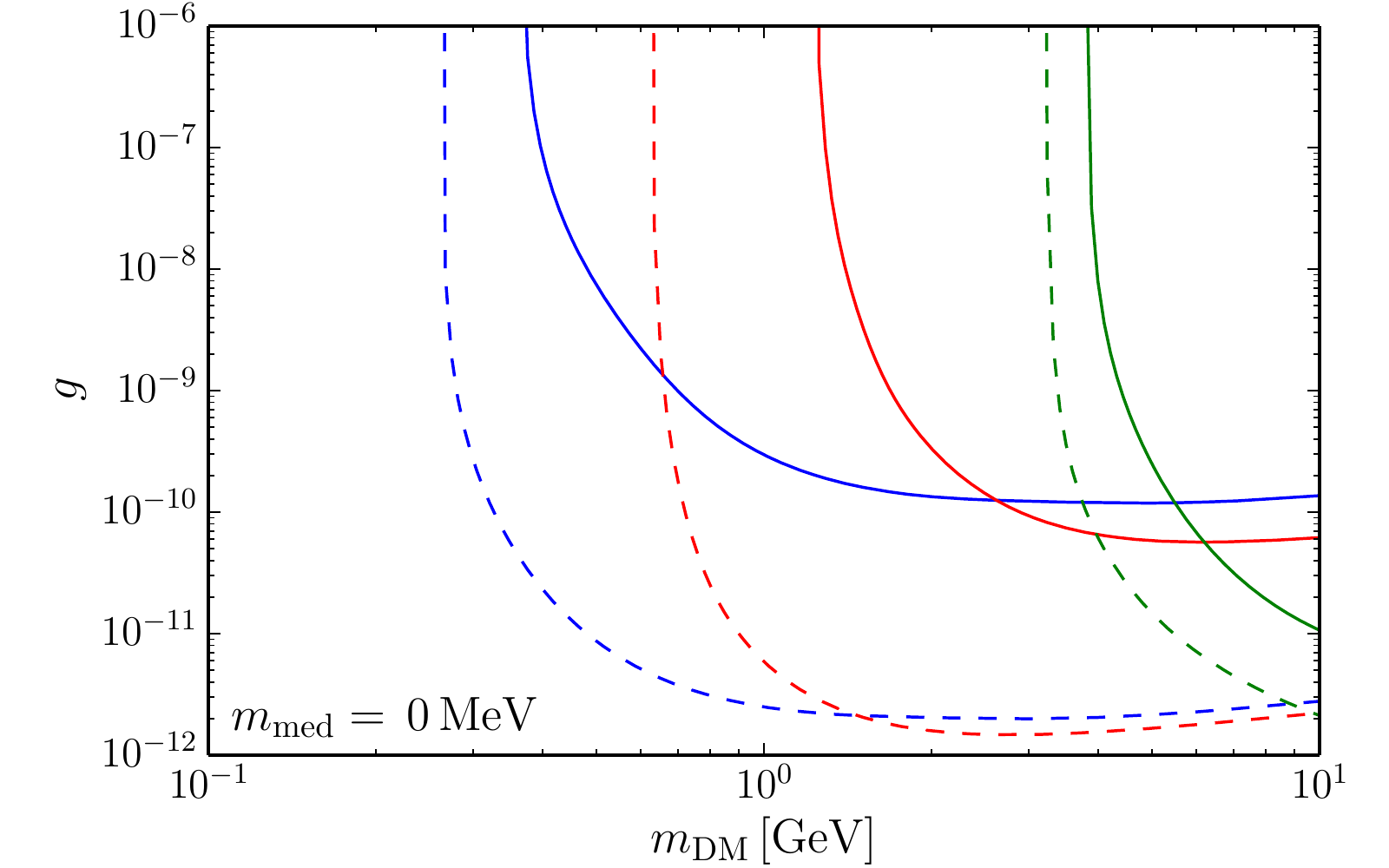}
\includegraphics[width=0.48\textwidth]{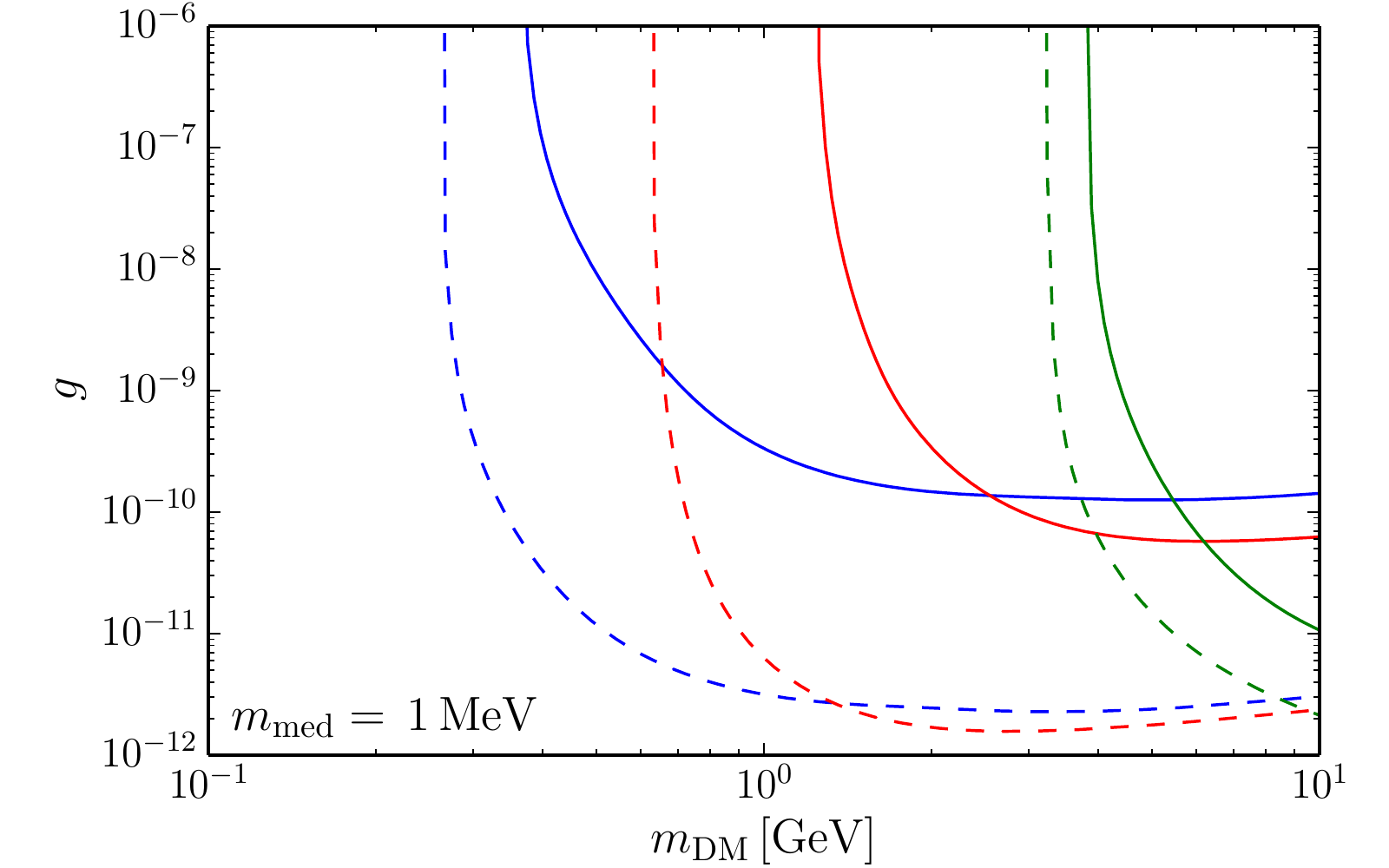}\\
\includegraphics[width=0.48\textwidth]{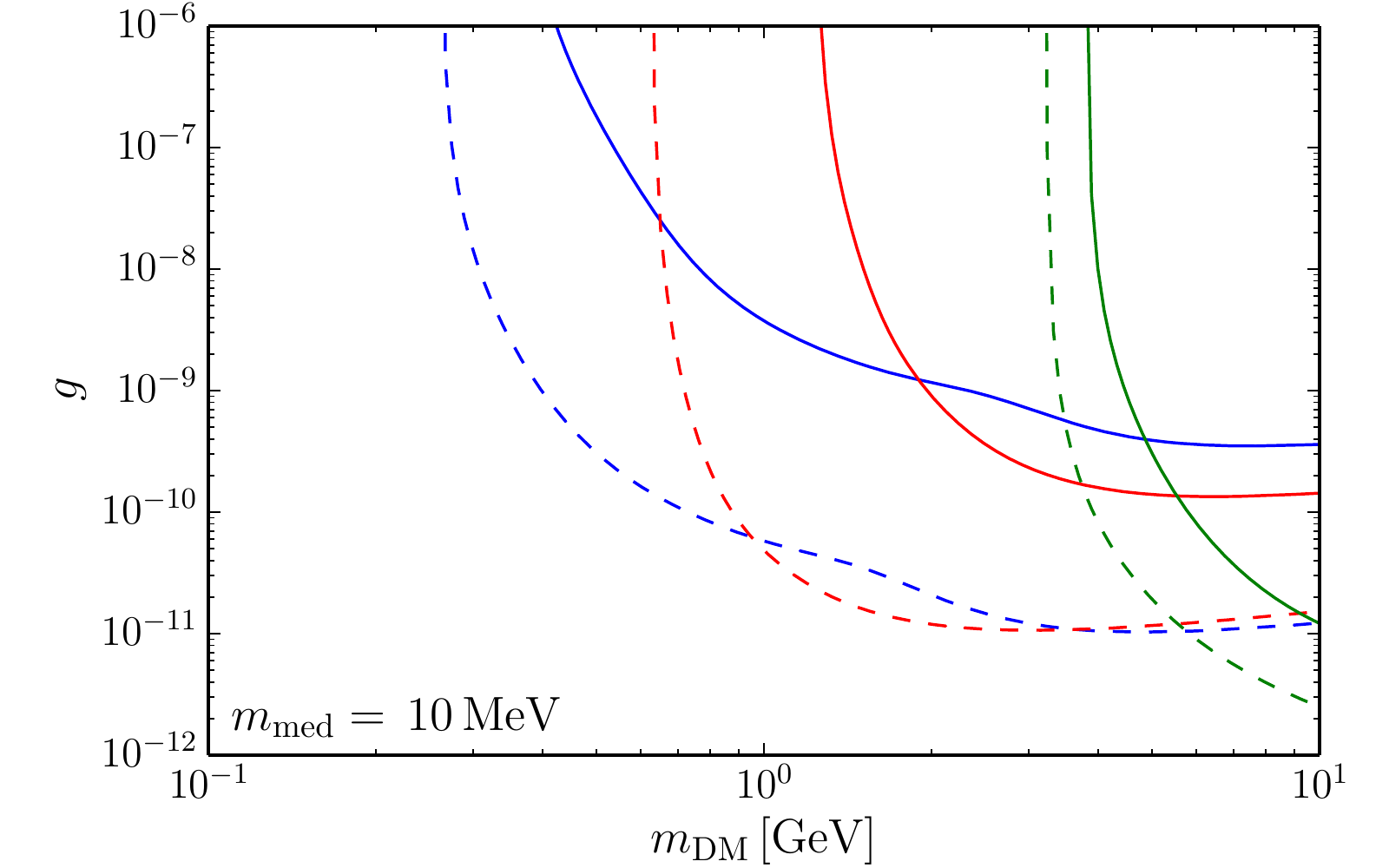}
\includegraphics[width=0.48\textwidth]{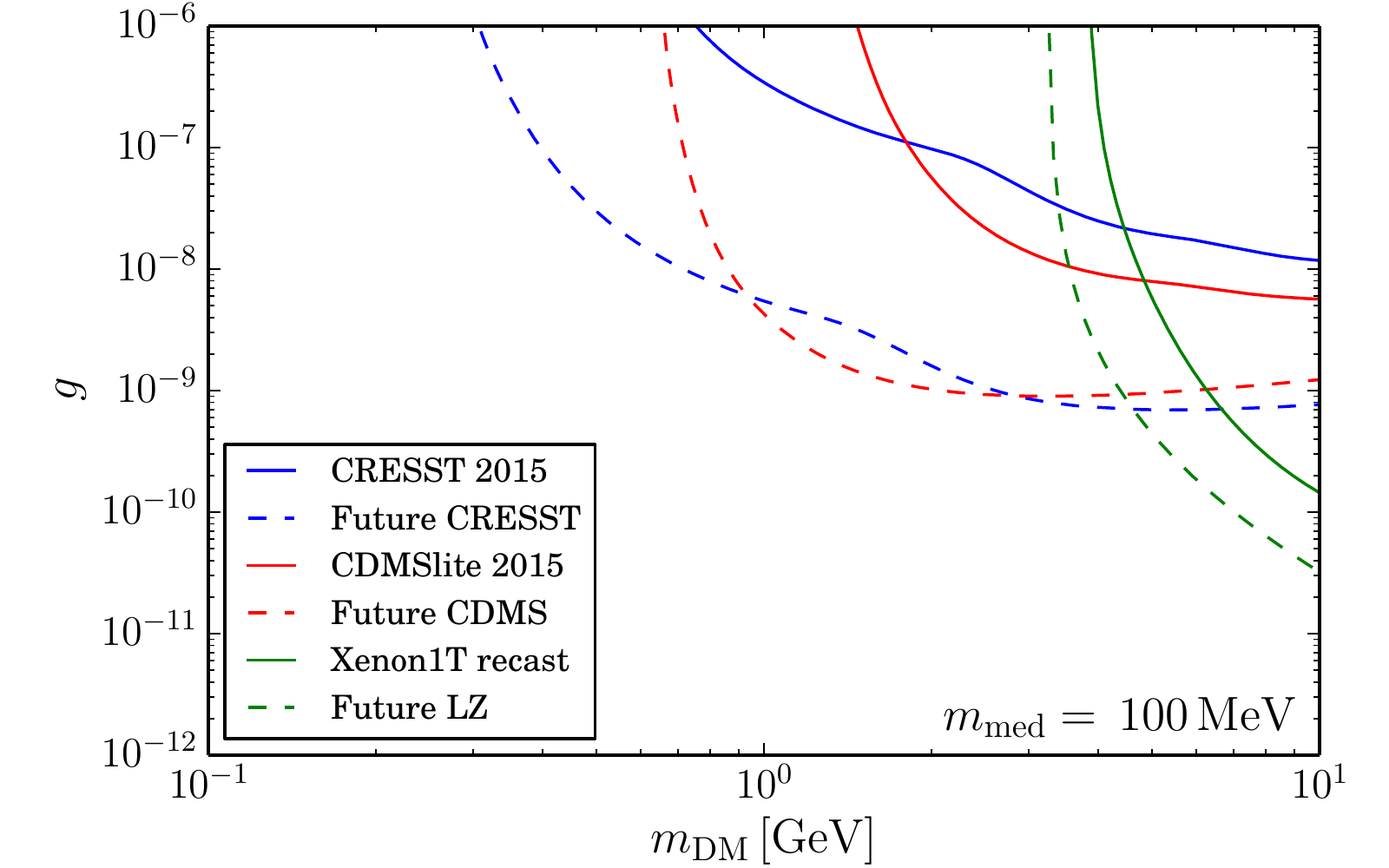}
\caption{\label{fig:mediator}Constraints (at 90\% CL) on the effective coupling $g$ for DM interacting via a light mediator. These bounds assume that the mediator couples only to protons ($\theta = 0$).}
\end{figure}

Having validated our implementation, we now show the corresponding constraints on the effective coupling $g$ for smaller mediator masses (figure~\ref{fig:mediator}). As expected, we observe that with decreasing mediator mass direct detection experiments become sensitive to smaller values of the effective coupling $g$, with particularly large enhancement factors found for low-threshold experiments. Another important observation is that for $m_\text{med} \ll q^\text{min}$ the shape of the recoil spectra (and hence the exclusion limit) becomes independent of $m_\text{med}$. We will show in more detail in section~\ref{sec:reconstruction} that the ability of direct detection experiments to reconstruct the mediator mass is limited to the range $q^\text{min} \lesssim m_\text{med} \lesssim q^\text{max}$.

We note that the constraints shown in figure~\ref{fig:mediator} depend on the assumed value of the coupling ratio $\theta$. Here we have chosen $\theta = 0$ corresponding for example to a vector mediator with kinetic mixing. In terms of the kinetic mixing parameter $\epsilon$ and the mediator-DM coupling $g_\text{DM}$ the effective coupling $g$ is then simply given by $g = e \, \epsilon \, g_\text{DM}$, where $e = \sqrt{4 \pi \alpha}$ is the electromagnetic coupling. Different values of $\theta$ would typically enhance the sensitivity of heavy targets like tungsten relative to light targets like oxygen, except for specific values of $\theta$ that lead to destructive interference between proton and neutron contributions (see section~\ref{sec:couplings}).

\section{Reconstructing particle physics parameters}
\label{sec:reconstruction}

From figure~\ref{fig:mediator} we make two central observations: first, if DM-nucleon scattering is due to the exchange of light mediators, cryogenic experiments will have the best sensitivity to such interactions up to DM masses of around $10\,\mathrm{GeV}$. And second, compared to current bounds this sensitivity will improve by up to two orders of magnitude in terms of the effective coupling $g$, corresponding to up to four orders of magnitude in terms of the scattering rate. These observations immediately raise the question what we can hope to learn from a DM signal in low-threshold direct detection experiments. In this section we will answer this question by generating mock data and using this data to perform a parameter reconstruction.

To determine the compatibility of a specific particle physics hypothesis (characterized by a set of parameters $\mathbf{x}$) with a given set of data, we construct a likelihood function $\mathcal{L}(\mathbf{x})$ as follows. For each individual experiment $\alpha$, we calculate the Poisson likelihood
\begin{equation}
-2 \log \mathcal{L}^\alpha(\mathbf{x}, \mathbf{y}) = 2 \sum_i \left[ R^\alpha_i(\mathbf{x}, \mathbf{y}) + B^\alpha_i(\mathbf{y}) - N^\alpha_i + N^\alpha_i \log \frac{N^\alpha_i}{R^\alpha_i(\mathbf{x}, \mathbf{y}) + B^\alpha_i(\mathbf{y})} \right] \; ,
\label{eq:binnedL}
\end{equation}
where the sum is over all bins, and $R^\alpha_i$, $B^\alpha_i$, and $N^\alpha_i$ denote the number of predicted signal events, predicted background events and observed events, respectively. In addition to the particle physics parameters $\mathbf{x}$ we have introduced a number of nuisance parameters $\mathbf{y}$, which represent for example astrophysical or experimental uncertainties and may affect both signal and background predictions. These nuisance parameters may be constrained by an additional likelihood function $\mathcal{L}^\text{n}$. The total profile likelihood is then given by the product of the individual likelihoods maximised with respect to the nuisance parameters:
\begin{equation}
 \mathcal{L}(\mathbf{x}) = \max_\mathbf{y} \mathcal{L}^\text{n}(\mathbf{y}) \prod_\alpha \mathcal{L}^\alpha(\mathbf{x}, \mathbf{y}) \; .
\end{equation}

For the purpose of parameter estimation, the next step is to determine the value of the particle physics parameters $\mathbf{x}$ that maximise the profile likelihood, called $\mathbf{x}_0$. We can then construct the likelihood ratio $\mathcal{R}(\mathbf{x}) = \mathcal{L}(\mathbf{x}) / \mathcal{L}(\mathbf{x_0})$, which by definition is smaller than unity. Under random fluctuations in the data, the quantity $-2 \log \mathcal{R}(\mathbf{x})$ is expected to follow a $\chi^2$ distribution with number of degrees of freedom $n$ equal to the number of parameters $\mathbf{x}$. We can therefore exclude a hypothetical set of parameters $\mathbf{x}$ at confidence level $1-p$ if
\begin{equation}
1 - \text{CDF}_{\chi^2}(n, -2 \log \mathcal{R}(\mathbf{x})) < p \; ,
\end{equation}
where $\text{CDF}_{\chi^2}(n, x)$ denotes the cumulative distribution function for the $\chi^2$ distribution with $n$ degrees of freedom. For the case of two parameters, the $95\%$ confidence level (CL) bound is therefore given by $-2 \log \mathcal{R} < 5.99$.

In the following we will focus on $m_\text{DM} \lesssim 5\,\text{GeV}$, where cryogenic detectors have better sensitivity than liquid xenon experiments (see figure~\ref{fig:mediator}). We will first focus on one specific benchmark case, namely $m_\text{DM} = 2\,\text{GeV}$, $m_\text{med} = 3\,\text{MeV}$ and $\theta = 0$, and then discuss alternative benchmarks in section~\ref{sec:scenarios}. The assumed value of $g$ is chosen to be compatible with existing direct detection constraints. Choosing $g$ close to current exclusion limits will lead to an optimistic scenario, in which thousands of events can be observed in future experiments, whereas smaller values of $g$ imply smaller statistics and less precise parameter reconstruction. In the following, we will consider two alternative choices, namely $g = 2 \cdot 10^{-11}$ (referred to as the low-statistics case) and $g = 6 \cdot 10^{-11}$ (the high-statistics case). For our benchmark scenario, these choices correspond to around $900$ and $8000$ events across the set of future experiments that we consider (with SuperCDMS SNOLAB predicted to observe about four times as many events as CRESST-III).

We can now generate mock data for our benchmark scenario and the two possible coupling choices and then determine which alternative choices of $m_\text{DM}$ and $m_\text{med}$ are compatible with this data. For the purpose of parameter reconstruction it is sufficient to consider mock data sets without Poisson fluctuations. Although in this case the best-fit point will have a very high likelihood, $\mathcal{L} \approx 1$, we nevertheless obtain reasonable estimates of the likelihood ratio $\mathcal{R}(\mathbf{x})$ expected in a typical realization of the experiments. We will return to the issue of Poisson fluctuations in the context of goodness-of-fit estimates in section~\ref{sec:gof}.

\begin{figure}
\centering
\includegraphics[width=0.46\textwidth]{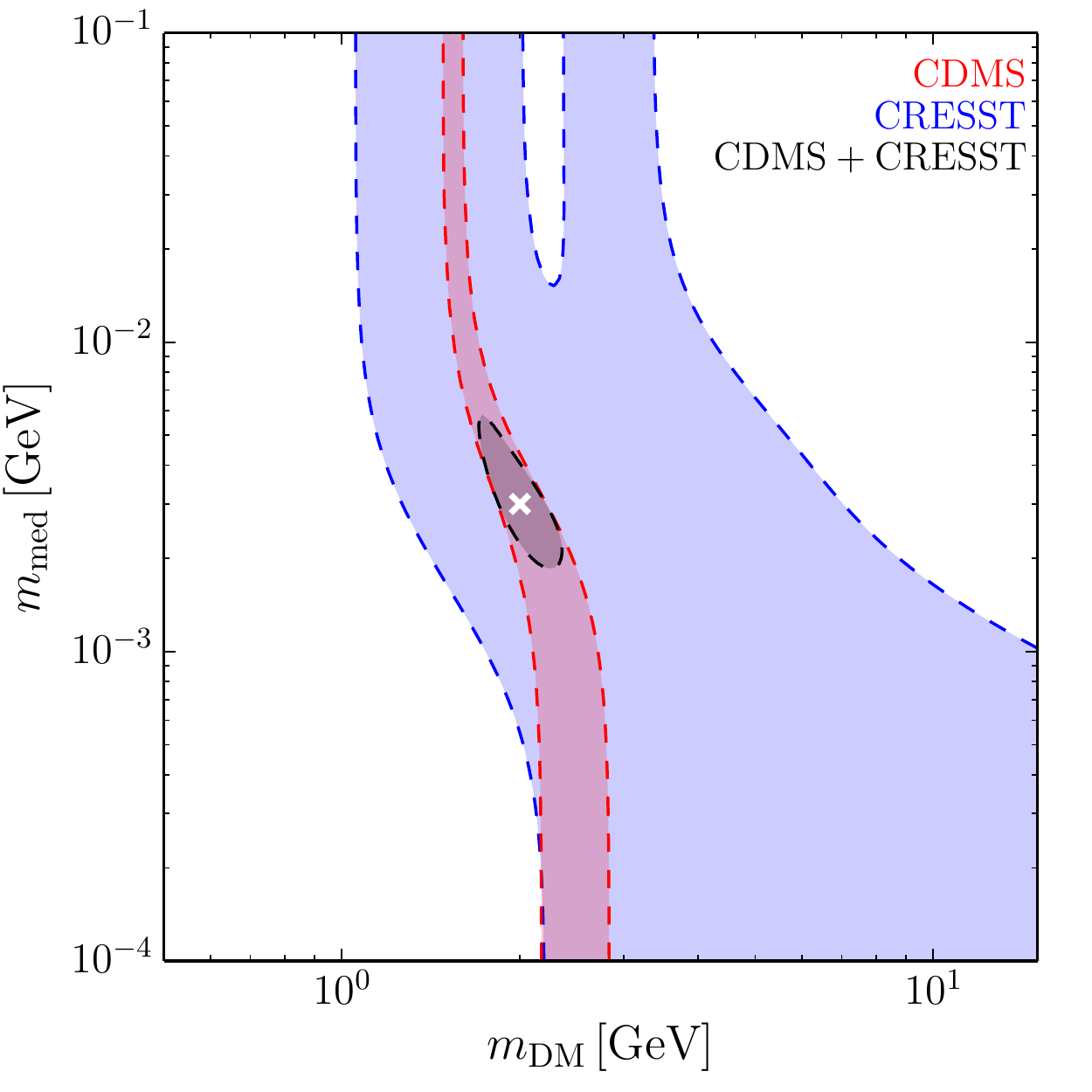}\hspace*{0.35cm}
\includegraphics[width=0.46\textwidth]{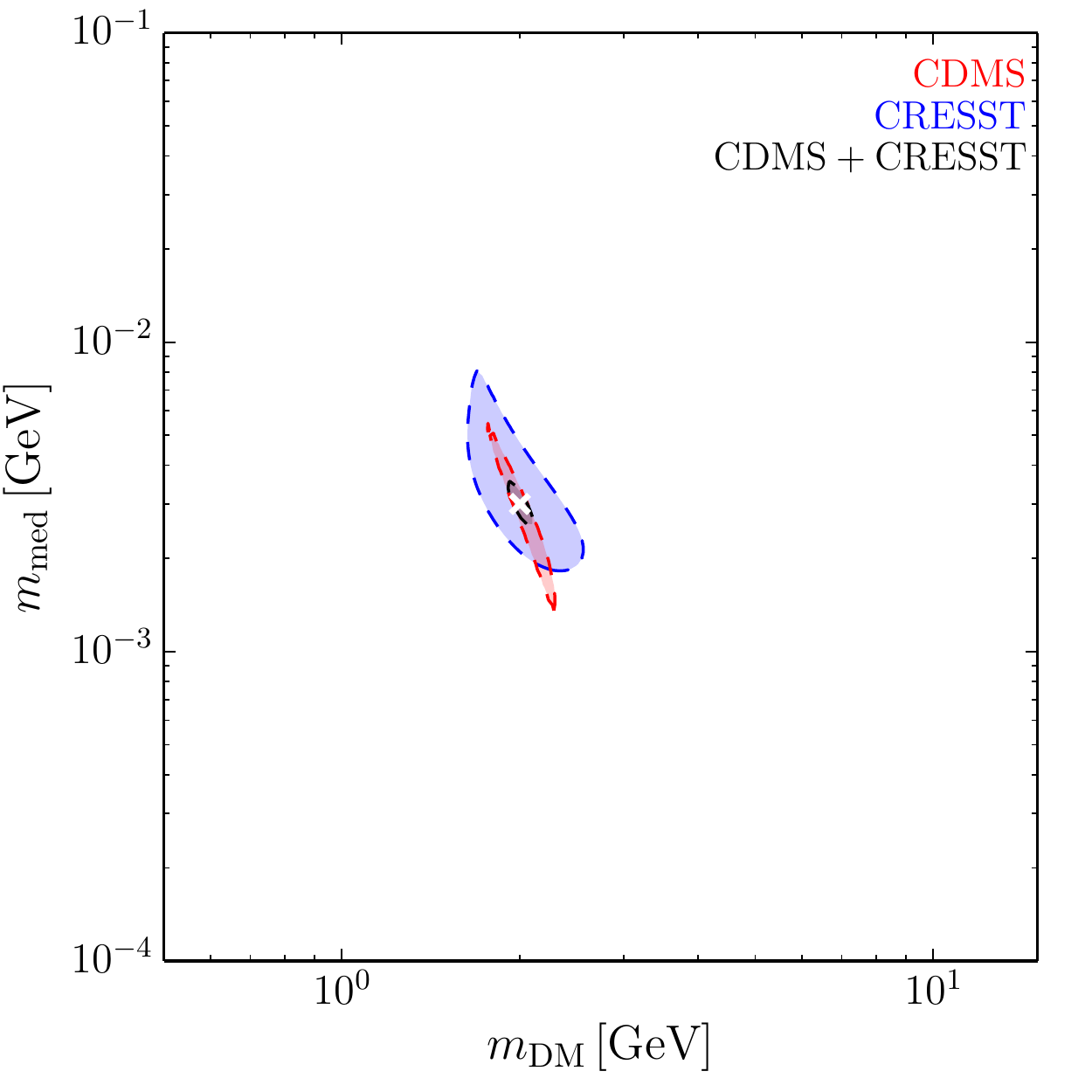}
\caption{\label{fig:reconst} Allowed parameter regions (at 95\% CL) reconstructed from a mock data set generated for $m_\text{DM} = 2\,\text{GeV}$, $m_\text{med} = 3\,\text{MeV}$ and $\theta = 0$ for the low-statistics scenario (left) and high-statistics scenario (right panel). For the purpose of reconstruction, we assume that $\theta$, the DM velocity distribution and the background normalization is known.}
\end{figure}

Figure~\ref{fig:reconst} shows the regions of parameter space compatible with the mock data generated for our benchmark scenario. For the purpose of these plots we are not interested in reconstructing the effective coupling $g$, i.e.\ we will simply treat it as a nuisance parameter and fix it to the value that maximises the likelihood. Nevertheless, the assumed value of $g$ does play an important role as it determines the number of events that we expect to observe. The left (right) panel corresponds to the low-statistics (high-statistics) case. Red and blue contours correspond to the parameter reconstruction based only on data from SuperCDMS SNOLAB and CRESST-III respectively, while the grey region indicates the combined constraints. Note that in these plots we do not yet take into account nuisance parameters related to background or astrophysical uncertainties; these will be discussed later in this section.

A striking feature in the left panel of figure~\ref{fig:reconst} is the accuracy of the parameter reconstruction from SuperCDMS SNOLAB as compared to CRESST-III. This happens because of two reasons: first, SuperCDMS SNOLAB is predicted to observe about four times more events than CRESST-III and hence has much better statistics. Second, several target elements contribute to the observed event rates in CRESST-III, leading to different ways in which a good fit to the observed data can be obtained. While for the benchmark case that we consider the event rate is dominated by scattering off oxygen (because tungsten recoils are below threshold), very similar recoil spectra are obtained for heavier masses and scattering off tungsten. This observation also explains the two different `branches' found for heavy mediator masses. With sufficient statistics it becomes possible to distinguish between the two possible scenarios and reject the solution with dominant scattering off tungsten (see right panel on figure~\ref{fig:reconst}).

Another interesting feature is that all reconstructed parameter regions exhibit a characteristic `tilt' in the sense that lighter mediators are necessary for fitting heavier DM masses and vice versa. The origin of this shape is that heavier DM masses predict flatter recoil spectra, while lighter mediators predict steeper recoil spectra. Increasing the DM mass while decreasing the mediator mass and the effective coupling $g$ may therefore leave the recoil spectra approximately unchanged. This degeneracy disappears once the mediator becomes so light (or so heavy) that direct detection experiments are effectively in the massless mediator limit (or the contact interaction limit). The recoil spectra then no longer depend on the precise value of the mediator mass.

Finally we make the crucial observation that combining data from SuperCDMS SNOLAB and CRESST-III allows for a much more precise reconstruction of the mediator mass than considering the individual experiments. The primary reason behind this is that the degeneracy between DM mass, mediator mass and effective coupling strength $g$ discussed above depends on the target element and therefore on the experiments (see eq.~(\ref{eq:dRdE})). This degeneracy is therefore effectively removed when combining data from several experiments. Nevertheless, it is of course conceivable that the degeneracy will reappear (or new degeneracies will arise) once we include various sources of uncertainties. We will therefore now discuss such uncertainties in detail and assess their impact on our results.

\subsection{Background uncertainties}
\label{sec:bg}

In the parameter reconstruction performed above we have assumed exact knowledge of the shape and normalization of the experimental background(s). This is an overly optimistic assumption and we will now discuss a more conservative approach. Of course, if we were to allow arbitrary background shapes and normalizations, any kind of DM signal could be absorbed into the background, making it impossible to claim anything but exclusion limits. Any parameter reconstruction therefore necessarily requires some knowledge on the distribution of backgrounds. Here we assume that the shape of the background in each experiment is known, but we keep the normalization completely free. In other words, we introduce a nuisance parameter $y^\alpha$ for each experiment $\alpha$ such that the background predictions in eq.~(\ref{eq:binnedL}) are given by $B^\alpha_i(\mathbf{y}) = y^\alpha B^\alpha_i$. We do not impose any constraints on the parameters $y^\alpha$ apart from the trivial requirement that they must be strictly positive. As explained in more detail in appendix~\ref{app:experiments}, we assume the backgrounds to be flat in energy both for CRESST-III and SuperCDMS SNOLAB. However, this assumption can easily be modified within our framework once more detailed informations about the future experiments are available.

\begin{figure}
\centering
\includegraphics[width=0.46\textwidth]{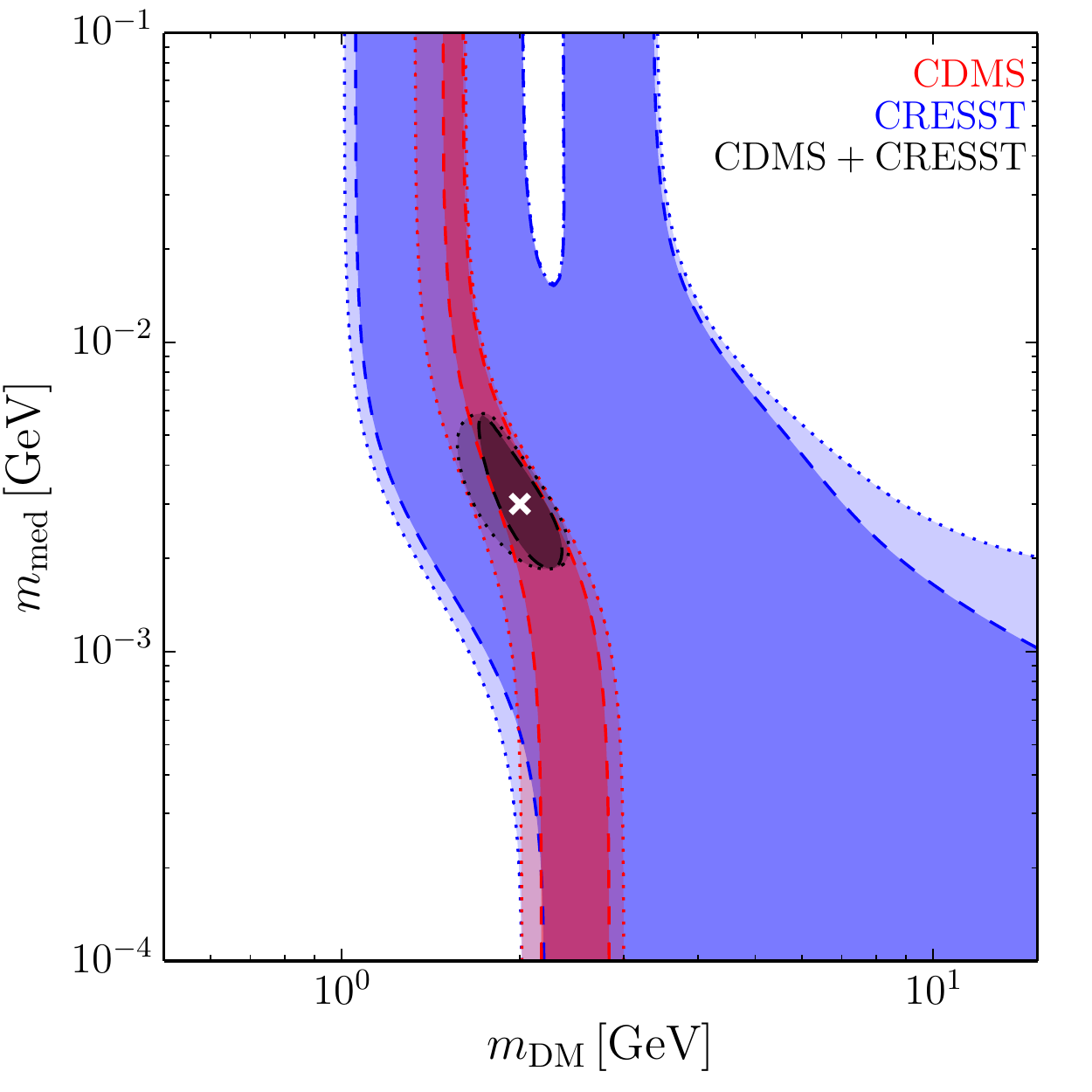}\hspace*{0.35cm}
\includegraphics[width=0.46\textwidth]{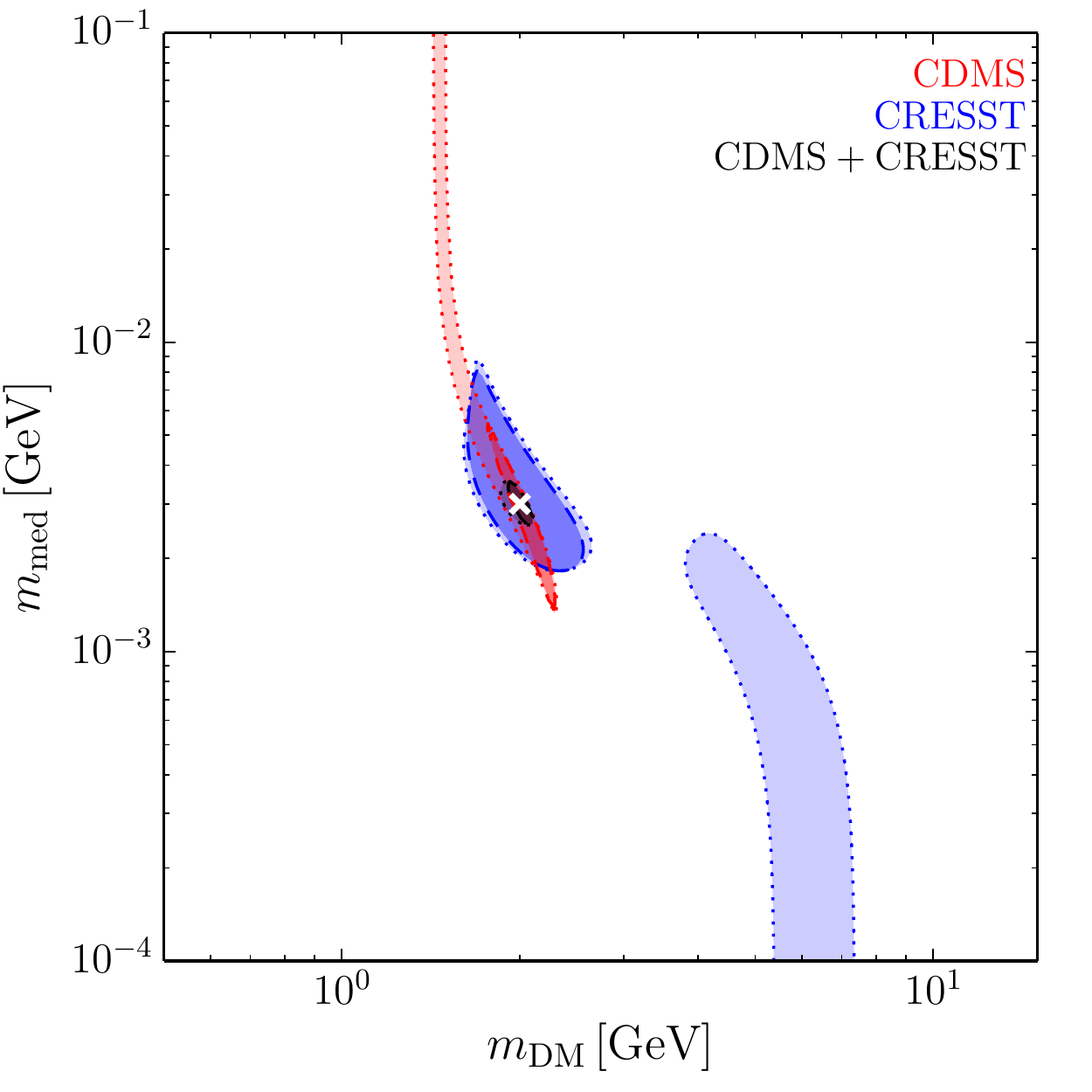}
\caption{\label{fig:reconst_bg} Same as figure~\ref{fig:reconst} but including additional nuisance parameters $y^\alpha$ to account for uncertainties in the background normalization (dotted curves). Dashed curves correspond to the fixed background normalization shown in figure~\ref{fig:reconst}.}
\end{figure}

We show the impact of including background uncertainties in figure~\ref{fig:reconst_bg}. As expected, these additional nuisance parameters visibly increase the size of the allowed parameter regions, in particular for SuperCDMS SNOLAB, where only a relatively small range of recoil energies is used to constrain the DM properties. For CRESST-III we observe that the second branch with scattering dominantly on tungsten now appears also in the high-statistics case. In principle it may be possible to distinguish these two branches, because the light yield of tungsten recoils in CRESST-III differs from the one for recoils on oxygen and calcium. Including this information (for example by constructing a two-dimensional likelihood in recoil energy and light yield) may hence make it possible to remove the second branch and obtain only one best-fit region. We leave this interesting possibility for future work.

In principle, given additional information on the different background contributions, it would be straight-forward to extend our approach and introduce individual nuisance parameters for each background source, such that not only the normalization of the background but also its shape can be varied. However, such detailed information is not presently available for the future projections that we consider. Nevertheless, we note in passing that the likelihood function defined in eq.~(\ref{eq:binnedL}) has an interesting property: as long as we only introduce nuisance parameters that rescale the signal prediction or (parts of) the background prediction, $-2 \log \mathcal{L}$ is a convex function of these parameters, so that any local minimum is necessarily a global one~\cite{Feldstein:2014ufa}. It is therefore numerically trivial to maximize the likelihood with respect to these nuisance parameters.

\subsection{Degeneracies with coupling ratios}
\label{sec:couplings}

So far we have made the assumption that only the mass of the mediator is unknown but its coupling structure is fixed. However, even in the simplest models of light mediators there are a number of different possible coupling structures~\cite{Frandsen:2011cg}. In fact, it is well-motivated to assume that the ratio of couplings to neutrons and protons is essentially a free parameter~\cite{Frandsen:2011cg}. If the effect of varying the mediator mass can be compensated by changing the coupling ratio, our lack of knowledge concerning the coupling structure of the mediator may affect our ability to determine its mass. In this section we discuss how the parameter reconstruction is affected if we do not make any assumptions on $\theta$.

For experiments consisting only of a single target element with charge number $Z_T$ and mass number $A_T$, the differential event rate is approximately proportional to\footnote{Strictly speaking, this relation only holds exactly for zero momentum transfer, because at finite momentum transfer there may be differences in the form factors for protons and neutrons~\cite{Zheng:2014nga}, but taking these effects into account is beyond the scope of this work.} $(Z_T \cos \theta + (A_T - Z_T) \sin \theta)^2$. This rescaling factor takes different values for different elements, so that varying $\theta$ will affect the combination of information from different experiments. In fact, varying $\theta$ can even change the shape of the recoil spectrum in a single experiment that consists of several different elements.

In contrast to the nuisance parameters introduced to parametrize the background uncertainties, the parameter $\theta$ enters into the likelihood in a more complicated way. The reason is that for $\theta < 0$ there will be destructive interference between proton and neutron contributions, which for a given experiment will be maximal if $\theta = \arctan(- Z_T / (A_T - Z_T))$. Consequently, there will generally be a number of different values of $\theta$ that maximize $\mathcal{L}(\theta)$ locally. We therefore perform an explicit scan over the full range $-\pi/2 < \theta < \pi / 2$ to determine the global maximum of $\mathcal{L}(\theta)$.

\begin{figure}
\centering
\includegraphics[width=0.65\textwidth]{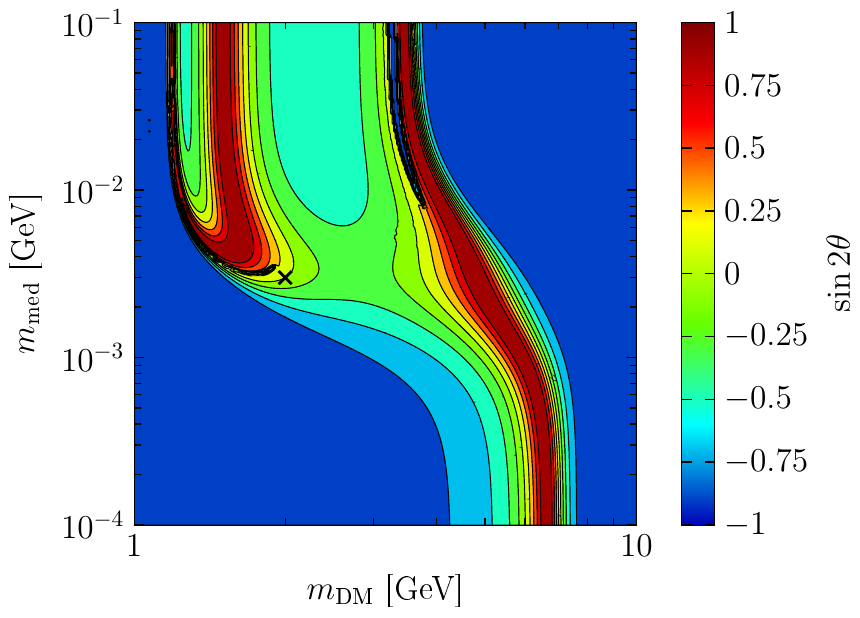}
\caption{\label{fig:theta} Best-fit values for the coupling ratio parameter $\theta$ as a function of the DM mass and mediator mass obtained from a set of mock data with the CRESST-III experiment. The mock data was generated under the assumption $m_\text{DM} = 2\,\mathrm{GeV}$, $m_\text{med} = 3\,\mathrm{MeV}$ and $\theta = 0$ (black cross).}
\end{figure}

If we consider SuperCDMS SNOLAB alone, the likelihood is essentially independent of $\theta$, because the recoil spectra for the different isotopes look so similar that a change in $\theta$ can almost entirely be compensated by a change in the effective coupling $g$. Indeed, we will see below that the allowed parameter regions for SuperCDMS SNOLAB remain essentially unchanged when including $\theta$ as a nuisance parameter. For CRESST-III, on the other hand, the situation is very different. Since the recoil spectrum for scattering of tungsten is typically much steeper than the recoil spectrum for scattering of oxygen and calcium, the shape of the recoil spectrum depends sensitively on $\theta$. Figure~\ref{fig:theta} shows the best-fit value of $\theta$ depending on the assumed values of $m_\text{DM}$ and $m_\text{med}$. We can make a number of interesting observations:
\begin{enumerate}
 \item The best-fit value for $\theta$ typically differs significantly from the value assumed to generate the mock data, covering almost the entire range $- \pi/2 \leq \theta \leq \pi/2$.
 \item If the assumed DM mass is large compared to the DM mass used to generate the mock data, the predicted recoil spectrum will be too flat, so the best fit can be obtained if scattering occurs exclusively on tungsten, i.e.\ for $\theta = -\pi/4$.
 \item If the assumed mediator mass is small compared to the mediator mass used to generate the mock data, the predicted recoil spectrum will be too steep, so the best fit can be obtained if scattering occurs exclusively on oxygen and calcium, i.e.\ for $\theta \sim -0.59$.
\end{enumerate}

\begin{figure}
\centering
\includegraphics[width=0.46\textwidth]{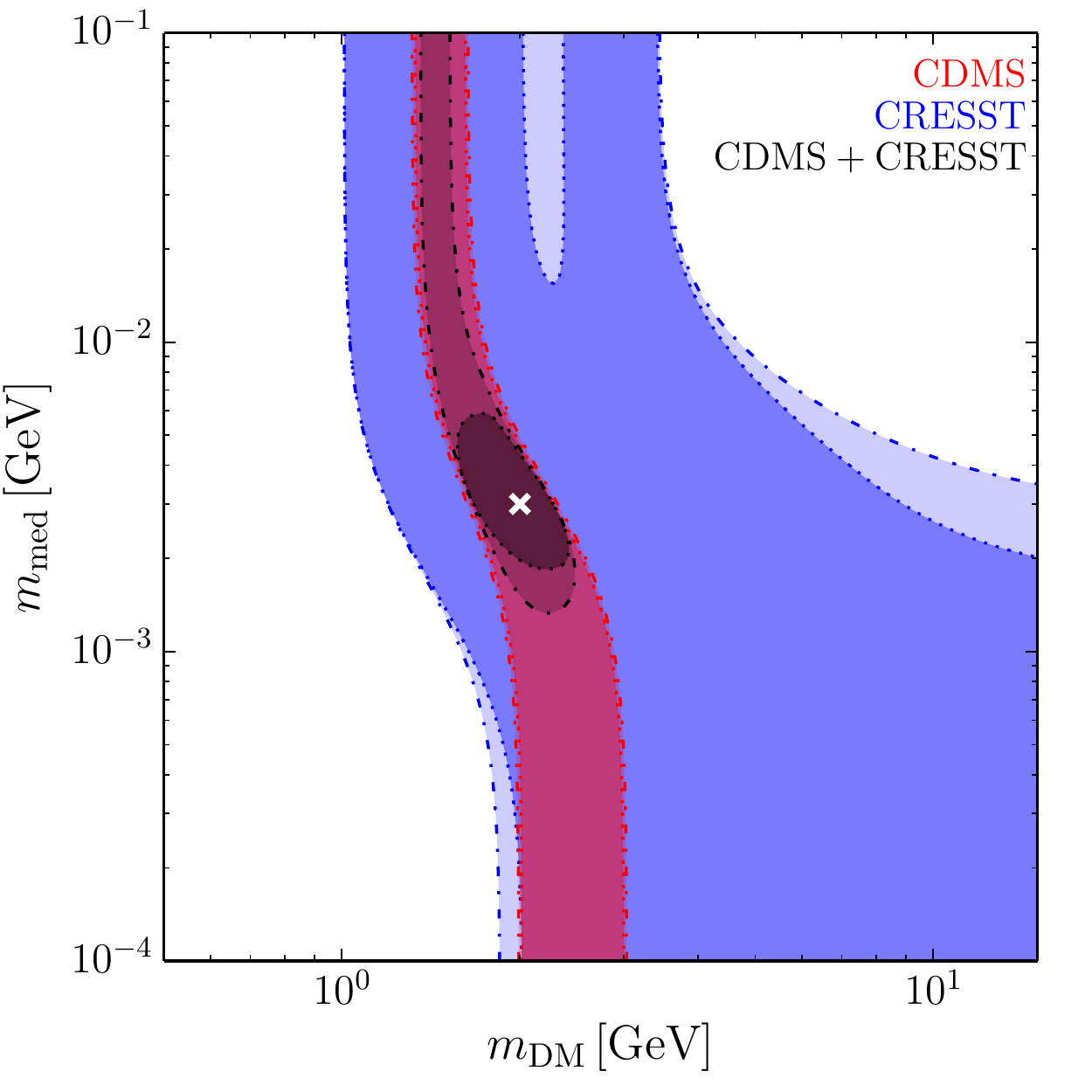}\hspace*{0.35cm}
\includegraphics[width=0.46\textwidth]{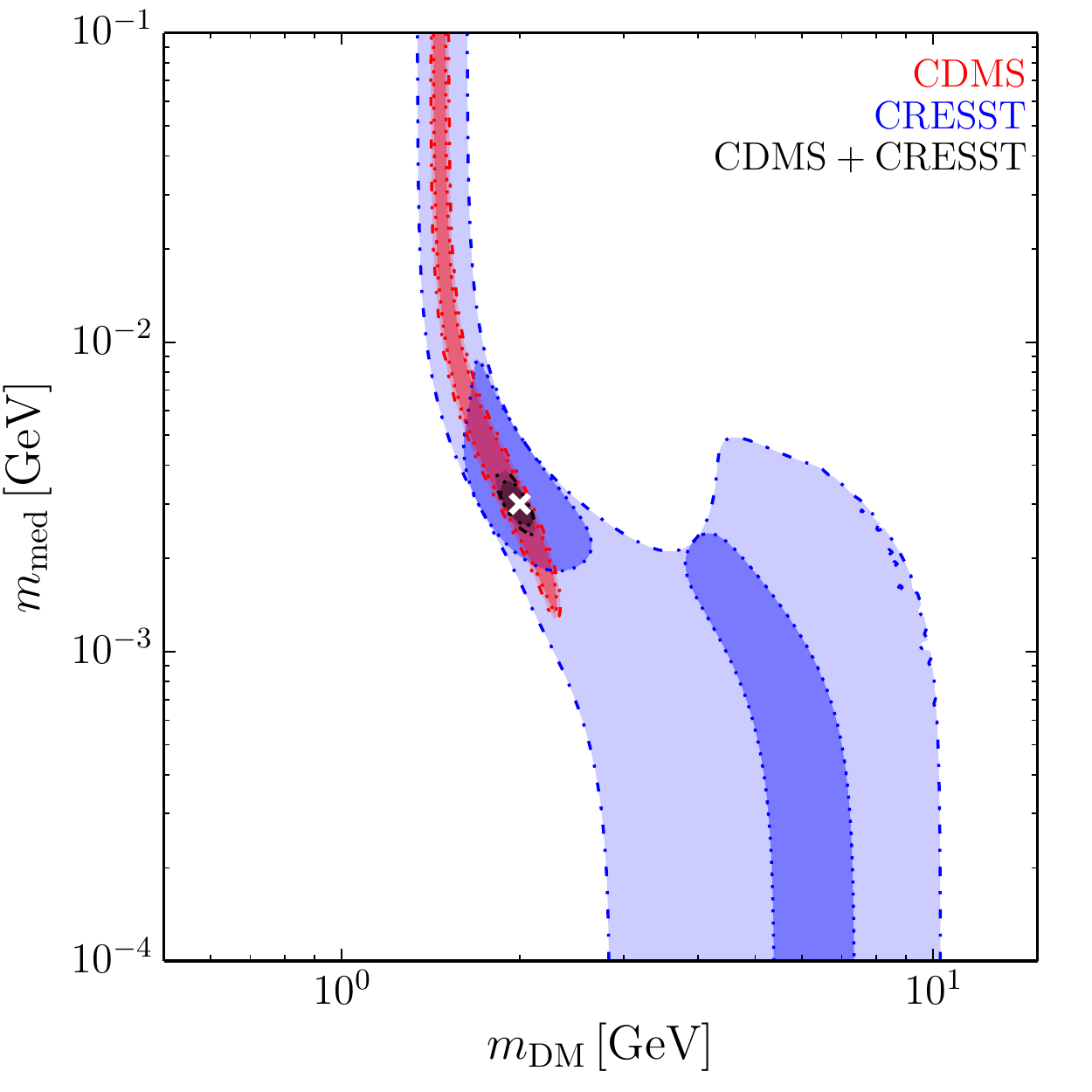}
\caption{\label{fig:reconst_theta} Same as figure~\ref{fig:reconst_bg} but including the unknown coupling ratio $\theta$ as an additional nuisance parameter (dash-dotted curves). The dotted curves correspond to $\theta = 0$, as already shown in figure~\ref{fig:reconst_bg}.}
\end{figure}

It should be clear from these observations that the ability of CRESST-III to reconstruct the mediator mass and the DM mass is significantly reduced when allowing for arbitrary values of $\theta$. This is confirmed explicitly in figure~\ref{fig:reconst_theta}. Clearly, in this case it is essential to combine the information from CRESST-III with data from SuperCDMS SNOLAB to perform a precise parameter reconstruction. In the absence of data from SuperCDMS SNOLAB it could also be interesting to attempt a discrimination between the different target elements in CRESST-III or to perform a combination of CRESST-III with the existing exclusion limits from Xenon1T, which would disfavour solutions with heavy DM scattering on tungsten.

\subsection{Astrophysical uncertainties}
\label{sec:astro}

There are two kinds of astrophysical uncertainties that may affect the interpretation of direct detection experiments: uncertainties in the local DM density $\rho_0$ and uncertainties in the DM velocity distribution $f(\mathbf{v})$. The local DM density $\rho_0$ enters linearly into the differential event rate, so varying this quantity is equivalent to varying the effective coupling $g$. Since we are not interested in determining $g$ and simply treat it as a nuisance parameter, our approach therefore already accounts also for uncertainties in the local DM density. The velocity distribution, on the other hand, enters in a more complicated way, giving rise to additional uncertainties that we will now discuss.

The differential event rate depends on the velocity distribution via the velocity integral $\eta(v_\text{min})$, which in turn depends on the DM mass and the recoil energy via $v_\text{min} = \sqrt{\frac{m_T \, E_\mathrm{R}}{2 \mu_{T}^2}}$. Changes in the velocity distribution may therefore change the shape of the recoil spectrum and thereby limit our ability to extract information on the particle physics parameters. One possible way to deal with astrophysical uncertainties are so-called halo-independent methods~\cite{Fox:2010bz,Frandsen:2011gi,Gondolo:2012rs}, which aim to combine information from different experiments in such a way that the dependence on $\eta(v_\text{min})$ drops out. This approach has been successfully applied to many different models and in particular to models with light mediators~\cite{Cherry:2014wia}.

However, as pointed out in ref.~\cite{Frandsen:2013cna}, there is a fundamental limitation of this approach in the case of low-mass DM. If $m_\text{DM} \ll m_T$ for all target elements under consideration, we find that $v_\text{min} \simeq \sqrt{\frac{m_T \, E_\mathrm{R}}{2 \, m_\text{DM}^2}} = \frac{q}{2 \, m_\text{DM}}$. The velocity integral hence depends on the same combination of $m_T$ and $E_\mathrm{R}$ that also enters in the factor for light mediator exchange, eq.~(\ref{eq:dRdE}). In other words, for low-mass DM, any change in the mediator mass can be compensated for by an appropriate change in the velocity integral for all target elements simultaneously. As a result, it will not be possible to infer \emph{any} information on the mediator mass if we allow for completely arbitrary velocity distributions.

We will therefore take a different approach and consider only velocity distributions of the Maxwell-Boltzmann form. This assumption is supported by recent studies involving numerical simulations of Milky Way like galaxies~\cite{Bozorgnia:2016ogo, Kelso:2016qqj}.  We can then study the impact of astrophysical parameters by varying the underlying parameters $v_0$, $v_\text{esc}$ and $v_\text{obs}$. In this case we can actually use the fact that $m_\text{DM} \ll m_T$ to our advantage. As shown in appendix~\ref{app:astro}, if we simultaneously rescale all three velocities by a common factor $z$, this change is fully equivalent to rescaling the DM mass by a factor $z$. We therefore introduce a new nuisance parameter $z$ and, rather than calculating the differential event rate as a function of $m_\text{DM}$, we calculate the differential event rate as a function of $z \, m_\text{DM}$. We restrict $z$ to lie in the range consistent with observations. At $95\%$ CL $v_0$ is constrained to lie in the range $\left[180 \mathrm{km \, s^{-1}}, 280 \mathrm{km \, s^{-1}}\right]$ while the range for $v_\text{esc}$ is approximately $\left[450 \mathrm{km \, s^{-1}}, 650 \mathrm{km \, s^{-1}}\right]$, see \cite{Green:2017odb} and references therein. These ranges can be reproduced if we require $0.8 \leq z \leq 1.2$ at $95\%$ CL. We implement this by means of a likelihood function for $z$ given by
\begin{equation}
\mathcal{L}^z = \frac{1}{\sqrt{2\pi} \sigma_z} \exp\left(- \frac{(z - 1)^2}{2 \, \sigma_z^2}\right)
\end{equation}
with $\sigma_z = 0.1$ and include this extra factor in the total likelihood.

\begin{figure}
\centering
\includegraphics[width=0.46\textwidth]{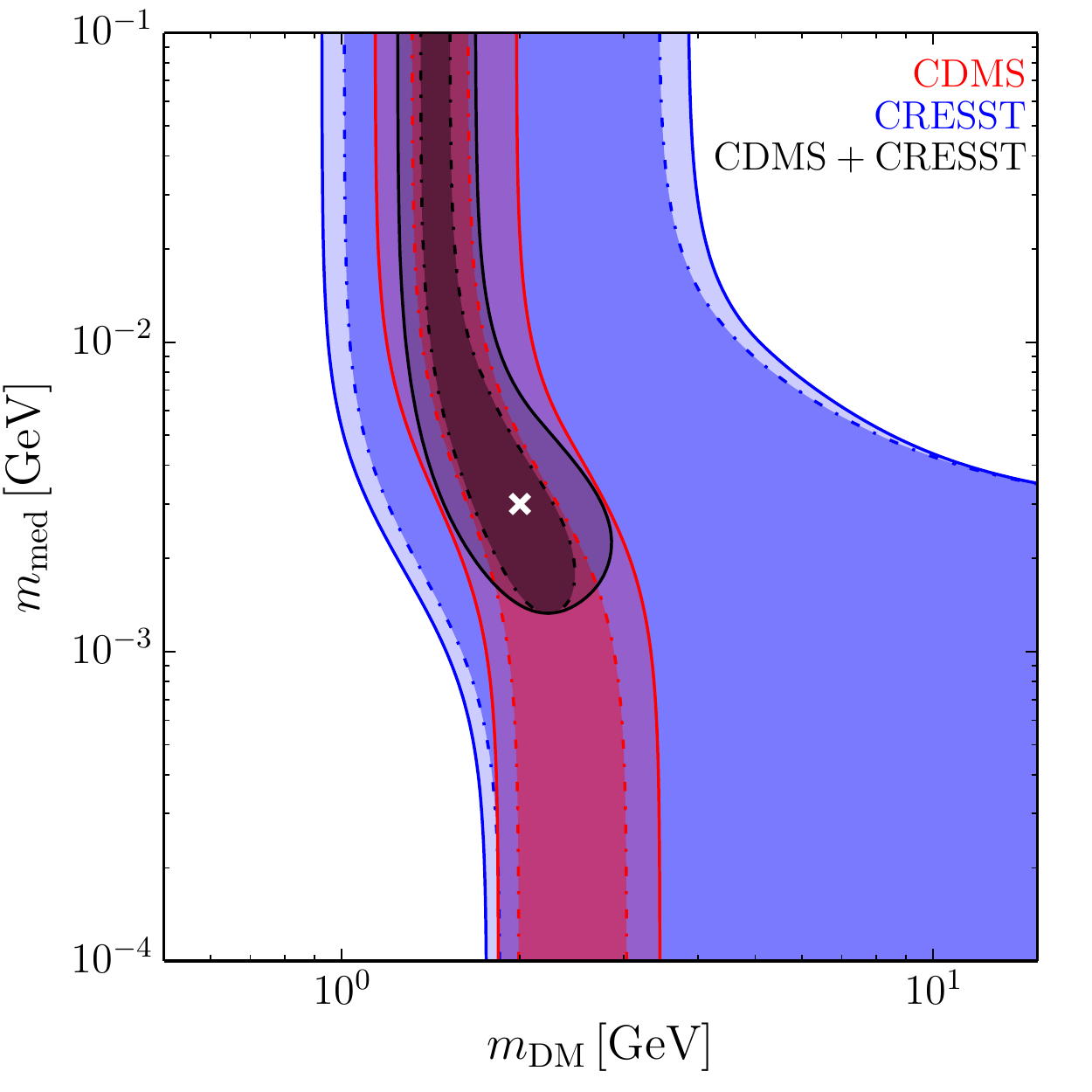}\hspace*{0.35cm}
\includegraphics[width=0.46\textwidth]{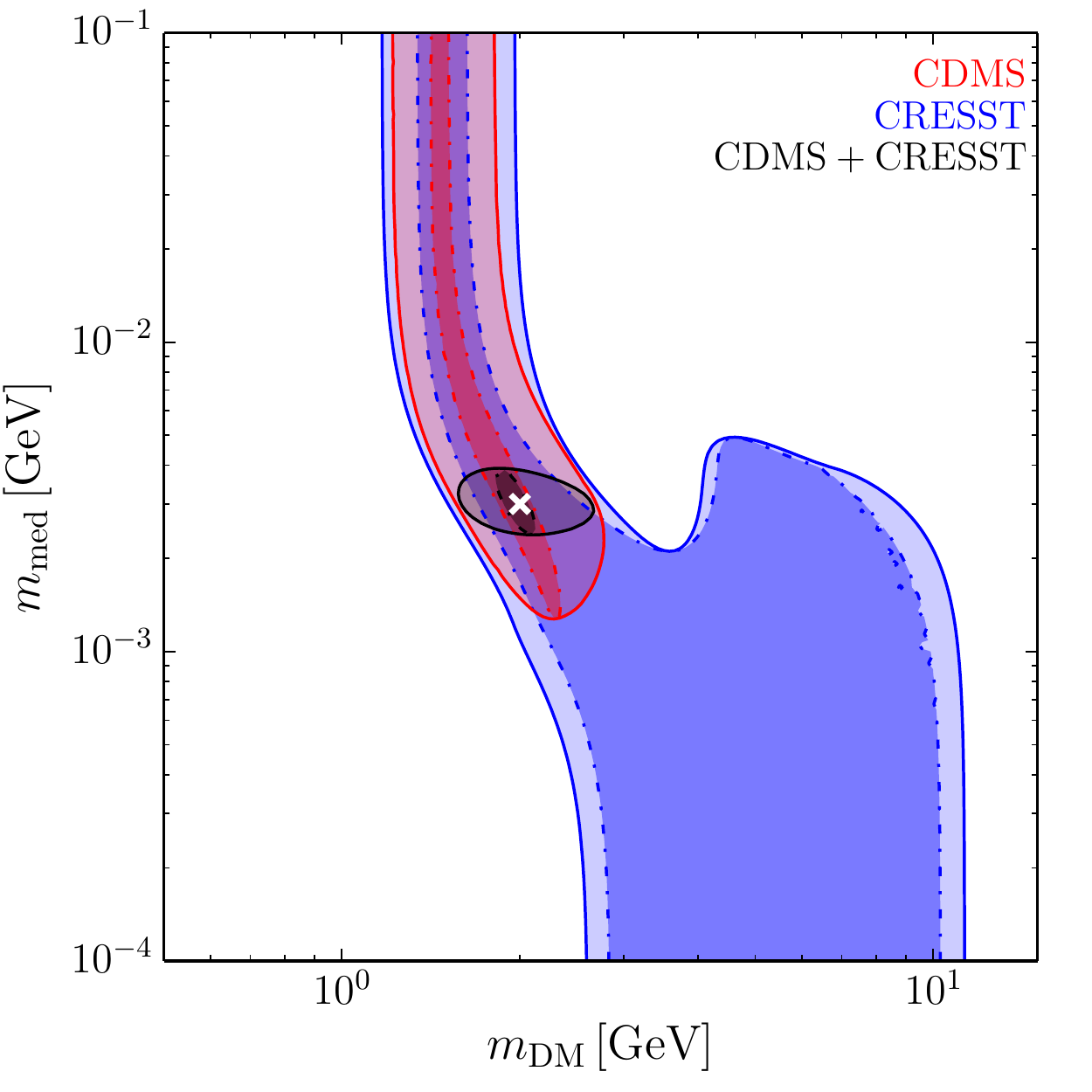}
\caption{\label{fig:reconst_astro} Same as figure~\ref{fig:reconst_theta} but including an additional nuisance parameter $z$ to parametrize astrophysical uncertainties (solid curves). See text for details. The dash-dotted curves correspond to a reconstruction not taking into account astrophysical uncertainties, as already shown in figure~\ref{fig:reconst_theta}.}
\end{figure}

As in the case of $\theta$ it is possible that the likelihood has several local maxima for different values of $z$, making it necessary to explicitly scan over all values of $z$ within the relevant range. Nevertheless, the simple way in which $\mathcal{L}$ depends on $z$ means that it is not in fact necessary to perform a two-dimensional scan over both $\theta$ and $z$, but rather that two separate one-dimensional scans are sufficient. Figure~\ref{fig:reconst_astro} shows the impact of including astrophysical uncertainties in addition to the uncertainties discussed above. As expected, the effect of varying $z$ is essentially to reduce our ability to reconstruct the DM mass, while not affecting our ability to reconstruct the mediator mass. Figure~\ref{fig:reconst_astro} constitutes our central result for the benchmark scenario: even when including a number of different nuisance parameters, an accurate reconstruction of the DM and mediator masses is possible given sufficient statistics.

\subsection{Alternative benchmark scenarios}
\label{sec:scenarios}

In the discussion above we have introduced a number of nuisance parameters that should be taken into account for a realistic assessment of the power of future low-threshold direct detection experiments. In addition to the two parameters that we are interested in reconstructing (the DM mass and the mediator mass), we have introduced two particle physics nuisance parameters (the coupling strength $g$ and the coupling ratio parameter $\theta$), one astrophysics nuisance parameter (the rescaling factor $z$) and one experimental nuisance parameter for each experiment (the background normalizations $y^\alpha$).\footnote{Even in the presence of these nuisance parameters it only takes a few seconds on a single CPU to calculate the profile likelihood for a single grid point and around an hour to perform the parameter reconstruction on a grid with $10^4$ points. A significant amount of computing time can be saved by reusing the information from neighbouring grid points to profile out the nuisance parameter related to astrophysical uncertainties.}  In this section we present our results for a number of different hypotheses on the particle physics properties of DM and discuss the physics interpretation.

\begin{figure}
\centering
\includegraphics[width=0.46\textwidth]{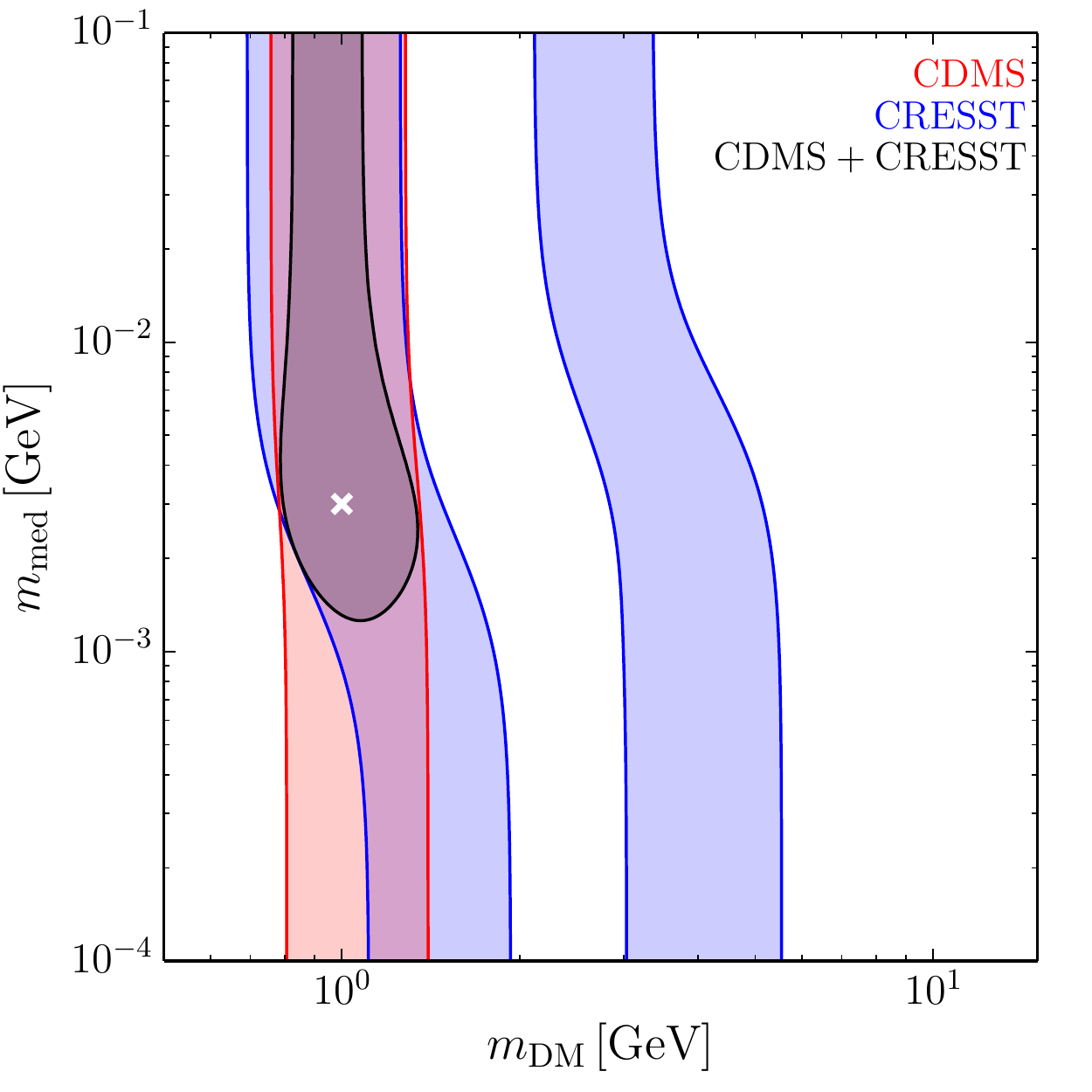}\hspace*{0.35cm}
\includegraphics[width=0.46\textwidth]{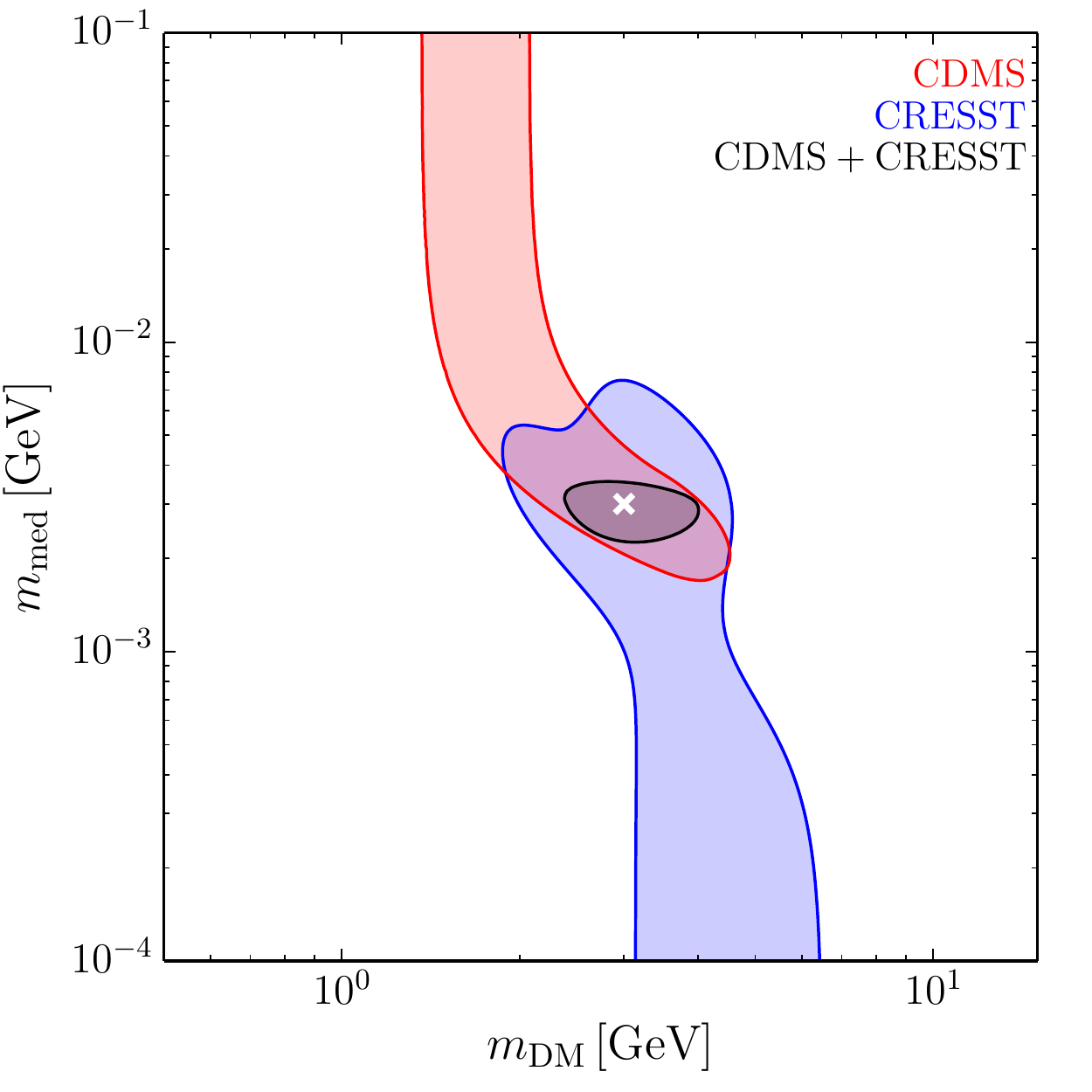}
\caption{Allowed parameter regions for two different DM masses. The mock datasets were generated assuming a DM mass of 1 (left) and 3 (right panel) GeV with a fixed mediator mass of $m_\text{med} = 3\,$MeV. We assume $g = 6 \cdot 10^{-11}$, which corresponds to different number of events in the left and right panel.}
\label{fig:HighStats_VaryingmDM}
\end{figure}

In figure~\ref{fig:HighStats_VaryingmDM} we perform a parameter reconstruction of the DM and mediator masses for two different assumptions on the true DM mass, namely 1 GeV (left panel) and 3  GeV (right panel). In both cases we fix the mediator mass to $m_\text{med} = 3\,$MeV and the effective coupling to $g = 6 \cdot 10^{-11}$.\footnote{It should be noted that this procedure leads to somewhat different numbers of events in the left and right panel. In particular the event rate in SuperCDMS is significantly suppressed for $m_\text{DM} = 1\,\text{GeV}$.} For $m_\text{DM} = 1\,\text{GeV}$ all observed events are very close to the low-energy threshold (i.e.\ within the first two or three bins). As a result the parameter reconstruction becomes more difficult and neither of the two experiments can individually constrain the mass of the mediator. For CRESST-III one furthermore finds a second branch of solutions corresponding to scattering off tungsten. Combining the information from both experiments leads to a somewhat better reconstruction, but the allowed parameter region still extends to arbitrarily heavy mediators. For heavier DM masses, on the other hand, an accurate parameter reconstruction is possible (see the right panel of figure~\ref{fig:HighStats_VaryingmDM}).

\begin{figure}
\centering
\includegraphics[width=0.46\textwidth]{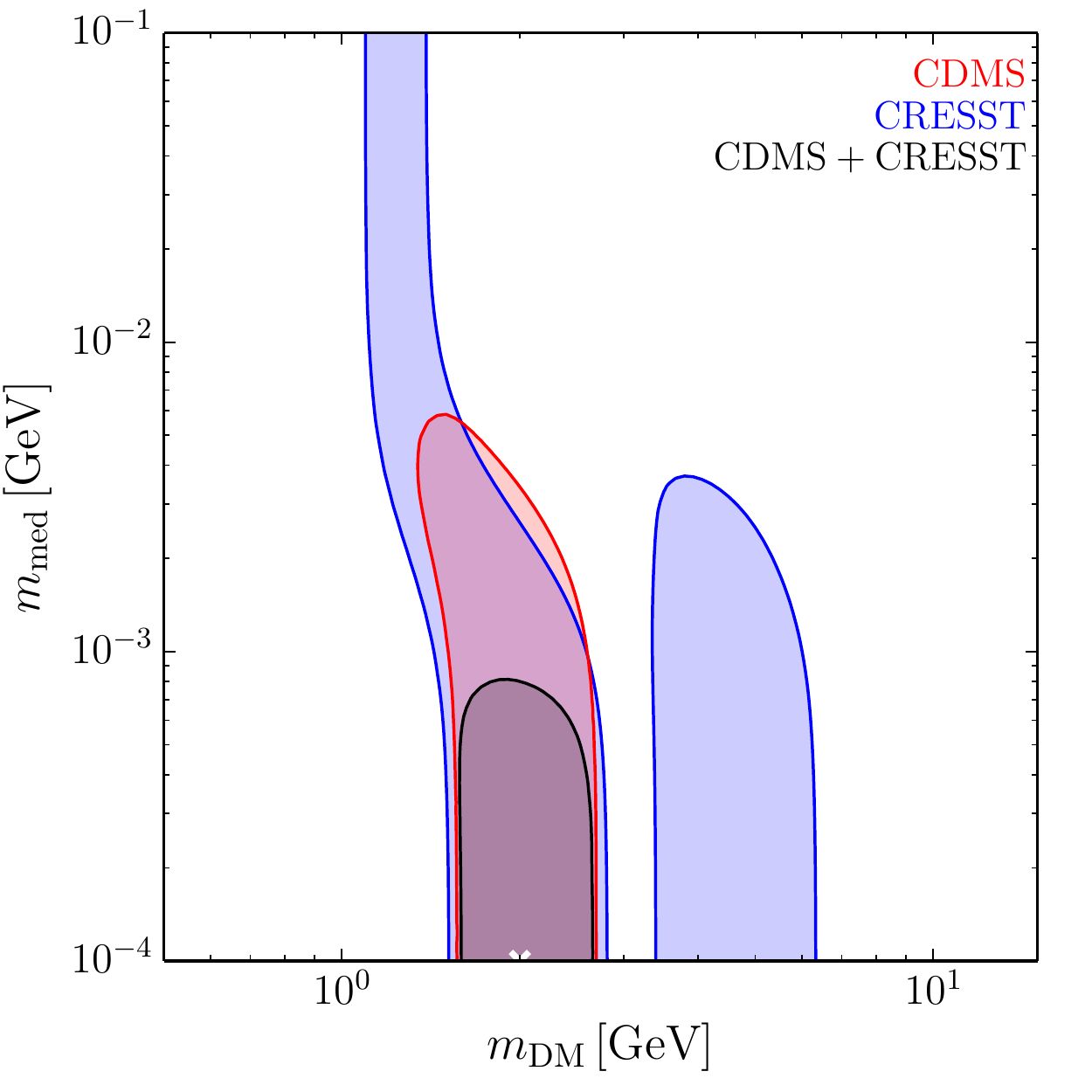}\hspace*{0.35cm}
\includegraphics[width=0.46\textwidth]{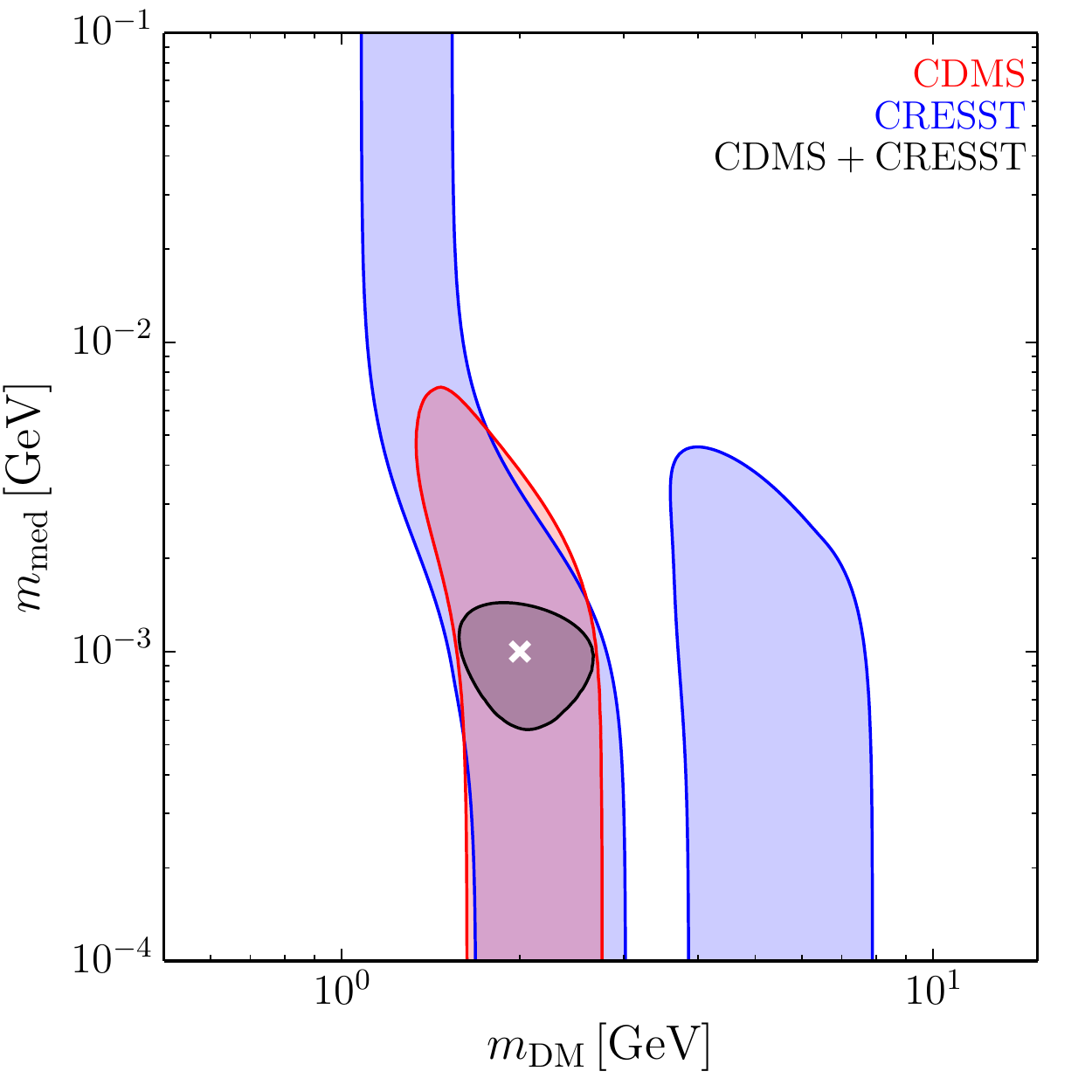}\\
\includegraphics[width=0.46\textwidth]{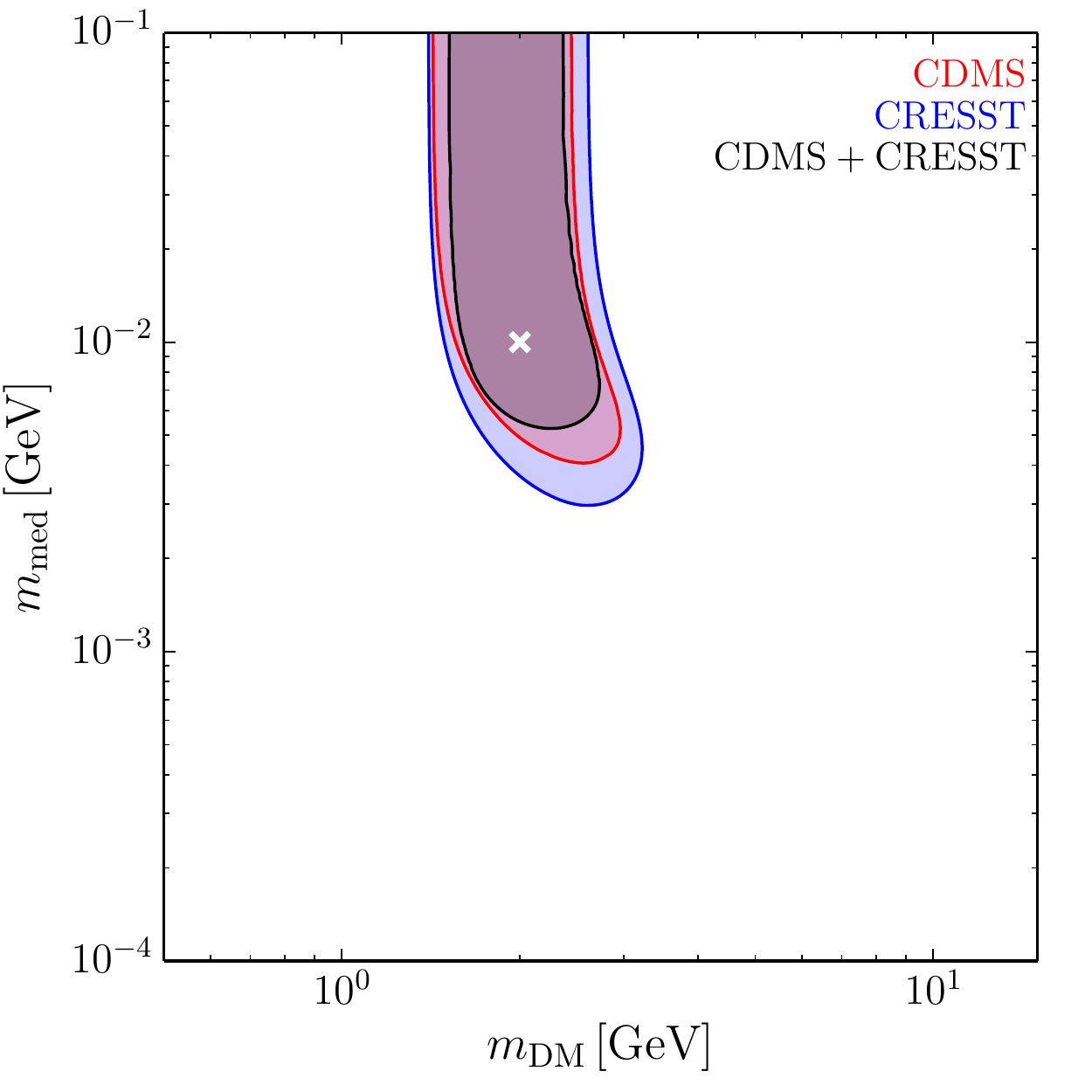}\hspace*{0.35cm}
\includegraphics[width=0.46\textwidth]{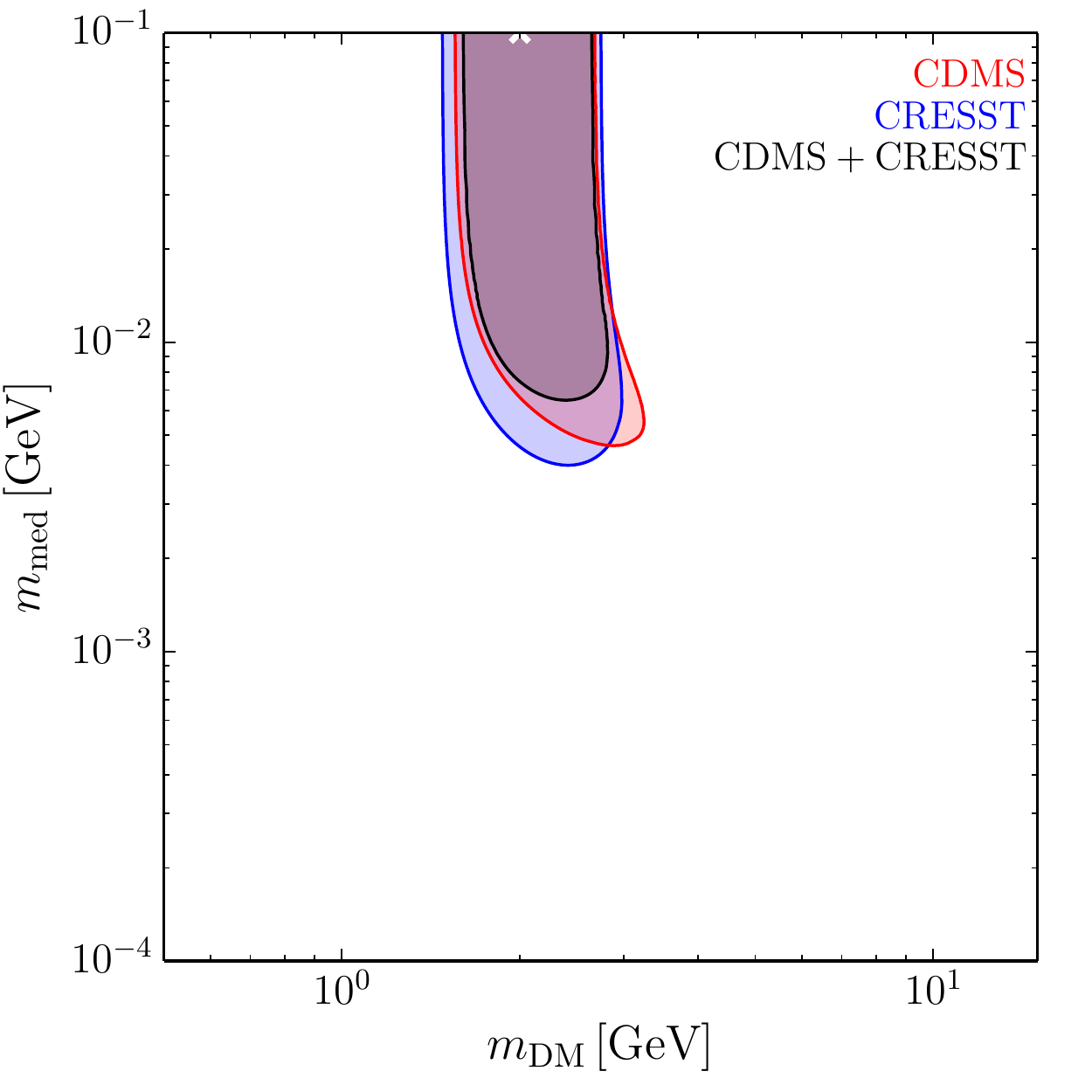}
\caption{Same as figure~\ref{fig:HighStats_VaryingmDM} but varying the true value of $m_\text{med}$. True value of $m_{\text{med}} = 0.1$ MeV (top left), 1 MeV (top right) 10 MeV (bottom left) and 100 MeV (bottom right) panel, keeping $m_\text{DM} = 2\,$GeV fixed. For each case, the true value of coupling $g$ is chosen such that the number of events corresponds to the high-statistics case discussed previously.}
\label{fig:HighStats_Varyingmmed}
\end{figure}

In figure~\ref{fig:HighStats_Varyingmmed} we investigate the effect of varying the assumed mediator mass while keeping the DM mass fixed to 2 GeV. For each mediator mass we choose the value of the coupling $g$ such that the predicted number of events is comparable to the high-statistics case discussed previously. In the top-left panel, the mediator mass is set to 0.1 MeV, i.e.\ effectively massless for the experiments under consideration. The combined fit to both experiments then places an upper bound on the mediator mass of about $m_\text{med} < 0.8 \, \text{MeV}$. As expected, the mock data is compatible with arbitrarily light mediators so that no lower bound can be placed. Conversely, if the assumed mediator mass is larger than about $10\,\text{MeV}$, it is no longer possible to distinguish our scenario from the case of contact interactions and the allowed parameter region extends up to arbitrarily high mediator masses (see bottom row of figure~\ref{fig:HighStats_Varyingmmed}). An accurate reconstruction of the mediator mass is possible only if the mass falls between these two extremes, as illustrated in the top-right panel for $m_\text{med} = 1\,\text{MeV}$.

To conclude this section we note that we have also studied the effect of making different assumptions on the value of $\theta$ used to generate mock data. If the effective coupling $g$ is adjusted in such a way that the event numbers are comparable to the ones discussed above, we find very similar results for different choices of $\theta$. An example with $\theta \neq 0$ will be discussed in section~\ref{sec:SIDM}.

\subsection{Goodness-of-fit estimates}
\label{sec:gof}

So far we have focussed on the issue of parameter estimation, i.e.\ we have constructed likelihood ratios to determine the parameter regions compatible with a given set of mock data. Another interesting topic that we can study in our framework are goodness-of-fit estimates, i.e.\ the question whether a specific choice of parameters yields a good description of the data. To answer this question we need to consider the absolute value of the likelihood rather than a likelihood ratio. Clearly, it is then no longer possible to neglect Poisson fluctuations in the data, because doing so would exaggerate the likelihood, i.e.\ would suggest unrealistically good agreement between data and model. In this section we therefore briefly discuss the effect of Poisson fluctuations and give a few examples for questions that can be answered with goodness-of-fit estimates.

In the limit of large bin counts the likelihood function defined in eq.~(\ref{eq:binnedL}) approaches a $\chi^2$ test statistic. We therefore expect that for the true parameters of nature $\mathbf{x}_0$ the likelihood $\mathcal{L}(\mathbf{x}_0)$ follows a $\chi^2$-distribution with $n = n_b - n_y$ degrees of freedom, where $n_b$ denotes the total number of bins across all experiments and $n_y$ denotes the total number of nuisance parameters that have been profiled out. For the two experiments that we consider $n_b = 29$, so if we profile out the effective coupling strength $g$ and the normalization of the background in both experiments, we expect to find $n = 26$. We confirm that this is indeed the case by performing a Monte Carlo (MC) simulation, i.e.\ by considering a large ensemble of mock data sets with Poisson fluctuations.

We can make use of this observation to study whether a given data set may enable us to exclude specific hypotheses about the particle physics nature of DM. For example, any DM signal observed in future direct detection experiments will first be interpreted under the assumption of a heavy mediator, i.e.\ contact interactions between DM and nuclei. A question of great interest would therefore be whether this hypothesis can be confidently excluded if the mediator is in fact light. To answer this question, we can use an ensemble of mock data sets generated under the assumption of a light mediator and calculate the likelihood under the \emph{incorrect} assumption of a heavy mediator. For this purpose, we treat both $m_\text{DM}$ and $\theta$ as nuisance parameters, i.e.\ we fix them to the value that maximizes the likelihood, such that $n = 24$.\footnote{Note that due to the degeneracy between the DM mass and the nuisance parameter $z$ parametrizing astrophysical uncertainties, including $z$ as an additional nuisance parameter would not increase the likelihood further and therefore does not reduce the relevant number of degrees of freedom.} For any specific mock data set, the heavy-mediator assumption can then be excluded at $95\%$ CL if $-2 \log \mathcal{L} > 36.4$.

Considering the same benchmark scenario as before ($m_\text{DM} = 2\,\text{GeV}$, $m_\text{med} = 3\,\text{MeV}$ and $\theta = 0$) we find that in the low-statistics case it is typically not possible to exclude the heavy-mediator hypothesis at $95\%$ CL, whereas in the high-statistics case a 95\%-CL exclusion is possible for about $98\%$ of the mock data sets. In the latter case it is also possible to exclude the hypothesis of a very light mediator (with $m_\text{med} \ll q^\text{min}$) at 95\% CL for more than 99.9\% of the samples.

\section{Connection to self-interacting dark matter}
\label{sec:SIDM}

The analysis performed in section~\ref{sec:reconstruction} applies to any model of DM interacting with nucleons via a light mediator, provided that the couplings of the mediator are spin-independent. In this section we apply these results to a particle physics model of particular interest, namely the case of self-interacting DM. 

It is well known that the presence of a light mediator can significantly enhance the rate of DM self-scattering in astrophysical systems, not only because of the long-range nature of the interactions but also because of additional non-perturbative effects due to multiple mediator exchange~\cite{Buckley:2009in, Loeb:2010gj, Aarssen:2012fx,Bellazzini:2013foa}. For a mediator with spin-independent interactions, these effects can be calculated by solving the non-relativistic Schroedinger equation for a Yukawa potential. The resulting scattering rate exhibits a characteristic velocity dependence such that the largest effects are expected on small scales, just as required in order to resolve the potential small-scale problems of collisionless cold DM~\cite{Weinberg:2013aya,Tulin:2017ara}. The quantity of interest for astrophysical observables is the momentum transfer cross section $\sigma_\mathrm{T}$, which should lie in the range $0.1 \, \mathrm{cm^2/g} \lesssim \sigma_\mathrm{T} / m_\text{DM} \lesssim 10 \, \mathrm{cm^2/g}$ in order to induce sizeable effects consistent with observations.

Out of the various possible particle physics realizations of this general scenario, the most attractive model consists of a fermionic DM particle $\psi$ (here assumed to be a Dirac fermion) and a scalar mediator $\phi$~\cite{Kaplinghat:2013yxa,Kouvaris:2014uoa}. For this model, DM self-annihilation is suppressed at small velocities due to CP conservation, evading the strong constraints from the Cosmic Microwave Background on many DM models with light mediators~\cite{Bringmann:2016din}. The effective coupling $g$ is then given by $g \approx 1.6 \times 10^{-3} \, y_\psi \, y_\text{SM}$, where $y_\psi$ denotes the DM-mediator coupling and $y_\text{SM}$ denotes the rescaled Yukawa coupling of the mediator to Standard Model (SM) fermions (i.e.\ the mixing of the scalar mediator with the SM Higgs boson).

A recent analysis of this model~\cite{Kahlhoefer:2017umn} concluded that for DM masses in the GeV region and mediator masses in the MeV region it is possible to reproduce the observed DM relic abundance and at the same time obtain large DM self-interactions if $y_\psi \sim 0.1$. Bounds from direct detection experiments then require $y_\text{SM} \lesssim 10^{-6}$ so that constraints from Higgs measurements and flavour physics are easily satisfied.\footnote{We note however that for such small values of $y_\text{SM}$ the lifetime of the mediator will be so large that it only decays after the onset of primordial nucleosynthesis. Nevertheless, the mass of the mediator is so small that it can only decay electromagnetically and with an energy small compared to typical nuclear binding energies. The dominant constraint is therefore expected to come from the entropy injected by the mediator decays, which modify the effective relativistic degrees of freedom $N_\text{eff}$. The magnitude of this effect depends however on additional aspects of the model, which are not relevant for the direct detection phenomenology. A detailed analysis of constraints from primordial nucleosynthesis is challenging and will be discussed elsewhere~\cite{Sebitobepublished}.} This mass range is precisely the parameter region that we have identified above to be the most interesting for cryogenic direct detection experiments. In fact, these experiments already place the strongest bounds on the SM-mediator coupling in this scenario. For the purpose of this section, we will therefore focus on scalar mediators and set $\theta = \pi/4$ as expected from Higgs mixing~\cite{Kaplinghat:2013yxa,Kouvaris:2014uoa}, both for the generation of mock data and for the subsequent parameter reconstruction.

\begin{figure}
\centering
\includegraphics[width=0.42\textwidth]{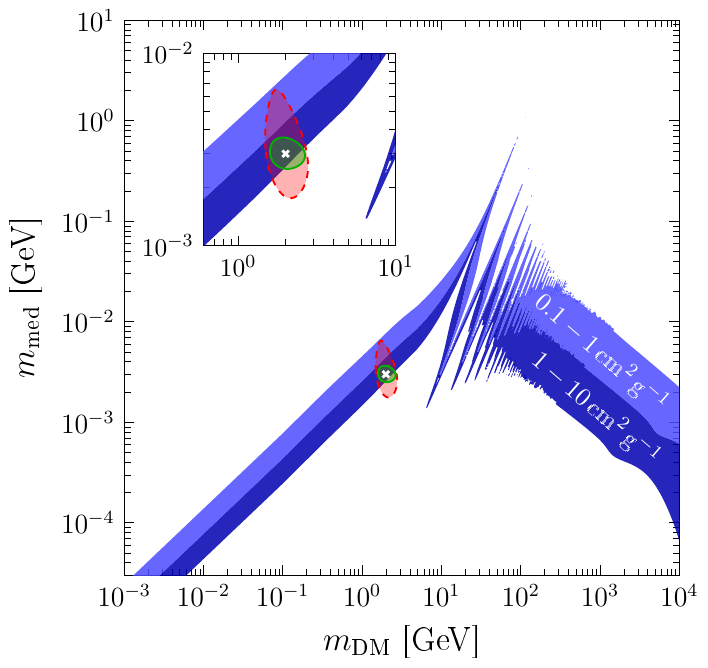} \qquad
\includegraphics[width=0.41\textwidth]{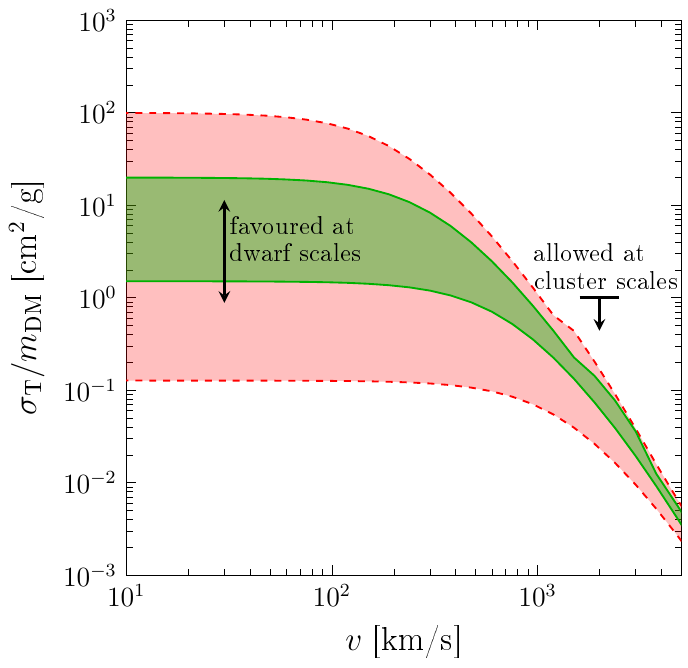}
\caption{\label{fig:SIDM} Left: reconstruction of the DM mass and the mediator mass compared to the parameter region that corresponds to large DM self-interactions on small astrophysical scales. Right: inferred DM momentum transfer cross section as a function of relative velocity. For both panels we have included all nuisance parameters discussed in section~\ref{sec:reconstruction}, except for the coupling ratio $\theta$ which we fix to $\pi/4$ (see text for details). Solid (dashed) contours correspond to high and low statistics scenario. Note that these figures assume that the DM relic abundance is set by thermal freeze-out via the process $\psi \bar{\psi} \to \phi \phi$.}
\end{figure}

While the DM momentum transfer cross section depends only on the DM-mediator coupling $y_\psi$, the DM-nucleon scattering cross section depends also on the mediator-SM coupling $y_\text{SM}$, which arises for example from mixing of $\phi$ with the SM Higgs boson. The effective coupling $g$ relevant for direct detection experiments can therefore be treated as an independent parameter in the same way as we have done above. Thus, if we can use direct detection experiments to infer the mediator mass and the DM mass, we can effectively determine the DM momentum transfer cross section (under the assumption that DM is a thermal relic~\cite{Bernal:2015ova}). This is illustrated in figure~\ref{fig:SIDM}. In the left panel we show a plot very similar to the ones shown in section~\ref{sec:reconstruction}, except that we also indicate the combinations of DM mass and mediator mass that lead to a momentum transfer cross section in the desired range~\cite{Kahlhoefer:2017umn}. In the right panel, we show the range of momentum transfer cross sections $\sigma_\mathrm{T}$ compatible with the inferred range of masses as a function of the DM relative velocity $v$. 

Since the DM self-interaction cross section depends very sensitively on the mediator mass, a precise determination of $\sigma_\mathrm{T}$ is clearly impossible. Nevertheless, figure~\ref{fig:SIDM} shows that it may be possible to demonstrate that both the magnitude and the velocity dependence of the DM self-scattering rates are broadly in agreement with astrophysical requirements, at least for the specific assumptions that we have made. Direct detection experiments may hence help us to determine not only the properties of the DM particle itself, but also to learn about its role in the formation of small-scale structures.

\section{Conclusions}
\label{sec:conclusions}

Light mediators communicating the interactions between DM and nuclei offer an interesting alternative to the effective operator approach conventionally adopted for the analysis of direct detection experiments. In this work we have demonstrated that cryogenic direct detection experiments are particularly well-suited for exploring this scenario. Their low threshold enhances the sensitivity to steeply falling recoil spectra and their excellent energy resolution allows for a precise reconstruction of the underlying particle physics from a potential signal. Present searches based on this technology already place stringent constraints on models with light mediators and significant improvements are expected for the next few years.

To illustrate the potential of future low-threshold detectors, we have performed parameter reconstructions from mock data sets in two planned experiments: SuperCDMS SNOLAB and CRESST-III. A strong emphasis has been placed on a realistic implementation of the experimental details. We include detector effects such as energy resolution, nuisance parameters for the background contributions as well as additional uncertainties reflecting both the unknown particle physics properties of DM and its astrophysical distribution. Even when including all of these uncertainties, low-threshold direct detection experiments maintain a remarkable potential to reconstruct the properties of the mediator.

For mediator masses in the range 1--10 MeV, the possibility to reconstruct mediator masses is mostly limited by statistics. Given sufficient exposure, it should be possible to reconstruct the mediator mass to better than a factor of 2. If the mediator is either lighter than about 1 MeV (and hence indistinguishable from truly long-range interactions) or heavier than about 10 MeV (and hence indistinguishable from contact interactions), a precise reconstruction of its mass is typically not possible. Nevertheless, with enough statistics one can clearly rule out the hypothesis of contact interactions for a very light mediator (and vice versa) using goodness-of-fit estimates.

The ability to reconstruct mediator masses relies strongly on the combination of different target materials. We have focussed on the combination of experiments based on CaWO$_4$ and germanium targets, but we have checked that a similarly good reconstruction can be obtained by combining data from a germanium experiment with results based on silicon detectors, which are being developed by the SuperCDMS collaboration. Another interesting, albeit challenging, possibility would be to exploit the differences in light yield for the different target elements in CRESST-III to statistically distinguish between models with dominant scattering on tungsten and those with dominant scattering on oxygen and calcium.

Throughout this work we have focussed on mediators with spin-independent interactions. Spin-dependent interactions are more challenging, since neither oxygen nor calcium nuclei carry spin and therefore the sensitivity of CRESST-III is significantly suppressed. It may be possible to gain complementary information with bubble chamber experiments like PICO-500~\cite{PICO}, but these lack the excellent energy resolution of cryogenic detectors. The most promising avenue appears to be the combination of germanium and silicon detectors, both of which possess at least some sensitivity to spin-dependent interactions. A detailed investigation of a wider range of interactions and a broader set of experimental technologies is left for future work.

Finally, we have pointed out that the range of mediator and DM masses that can be studied with cryogenic experiments coincides with parameter regions that have been considered in the context of self-interacting DM. This observation implies that, within specific model assumptions, one can translate between signals in direct detection experiments and astrophysical observables, such as core sizes in dwarf galaxies. By correlating these very different signatures, we can therefore hope to ultimately obtain a coherent picture of both the microscopic and macroscopic properties of DM.

\acknowledgments
We thank Achim G\"utlein for early collaboration and David Cerde\~no, Josef Jochum and Kai Schmidt-Hoberg for useful discussions. FK is supported by the
DFG Emmy Noether Grant No.\ KA 4662/1-1, SK is supported by the `New Frontiers' program of the Austrian Academy of Sciences and by FWF project number V592-N27, and SW is supported by the ERC Starting Grant `NewAve' (638528). SK thanks DESY for hospitality during the completion of this work.

\appendix
\section{Implementation of present and future experiments}
\label{app:experiments}

In this appendix we describe the experimental information and the assumptions that we make to construct the likelihood functions for the various experiments. 

\subsection*{CRESST-II}

The latest CRESST results~\cite{Angloher:2015ewa} are based on $52.2\,\mathrm{kg\,days}$ using the Lise detector module. We smear the physical recoil spectrum assuming a Gaussian energy resolution function with $\sigma = 62\,$eV~\cite{Angloher:2015ewa}, considering only fluctuations up to $3\sigma$ in order to avoid unphysical results far below threshold. The cut survival probabilities as a function of the detected recoil energy are taken from the recent data release~\cite{Angloher:2017zkf}. We then group the events observed in the acceptance region into 47 equidistant bins between 0.3 and 5.0 keV. For the calculation of the upper limits we conservatively set the number of expected background events to zero, which gives results similar to the optimum interval method~\cite{Yellin:2002xd} employed in ref.~\cite{Angloher:2015ewa}.

\subsection*{CRESST-III}

Our implementation of the future CRESST detector is based on the projections for the final state of CRESST-III presented in ref.~\cite{Angloher:2015eza}. The expected exposure is $1000\,\mathrm{kg\,days}$ with an energy threshold of $100\,$eV. We group the events in the energy range between 0.1 and $2\,$keV into 19 bins of width $0.1\,\mathrm{keV}$. Furthermore, we assume a background level of \mbox{$3.5\cdot10^{-2}\,\mathrm{keV^{-1}\,kg^{-1}\,day^{-1}}$}, corresponding to 3.5 events in each of the bins. The expected number of signal events is calculated assuming an efficiency of unity down to threshold, and using a Gaussian energy resolution of $20\,$eV. We take into account only recoils with a true energy above $60\,$eV in order to avoid unphysical upward fluctuations from very small recoil energies.

\subsection*{CDMSLite}

CDMSLite has analysed data from an exposure of $70.14\,\mathrm{kg\,days}$~\cite{Agnese:2015nto,CDMSlitedata}. We take the energy-dependent signal efficiency from~\cite{Agnese:2015nto} and follow the procedure outlined there to convert nuclear recoil energies (eVnr) into electron equivalent energies (eVee), using $k = 0.157$. We perform a fit to the width of various electron-capture peaks to determine the energy-dependent detector resolution. We then consider the energy range $[60\,\mathrm{eVee}, 500\,\mathrm{eVee}]$, which we divide into 10 bins of increasing size, such that the number of observed events in each bin is greater than one. Following the analysis in ref.~\cite{Agnese:2015nto}, we do not assume a background model, such that all observed events can potentially be DM signals. This procedure allows us to calculate a likelihood function that yields a bound very similar to the one obtained from the optimum interval method. 

\subsection*{SuperCDMS SNOLAB}

We estimate the sensitivity of SuperCDMS SNOLAB following ref.~\cite{Agnese:2016cpb}, focussing on the high-voltage germanium detectors. Although the SuperCDMS collaboration expects that a threshold as low as $40 \, \mathrm{eV}$ (nuclear recoil energy) can be achieved, we consider a low-energy threshold of $100\,\text{eV}$, so that it is a good approximation to assume the signal efficiency to be constant (at $85\%$) and the background level to be flat (at \mbox{$10\,\mathrm{keV^{-1}\,kg^{-1}\,year^{-1}}$})~\cite{Agnese:2016cpb}. We also limit ourselves to recoil energies below $300\,\text{eV}$ in order to avoid backgrounds from electron capture lines. This restriction will reduce the sensitivity of SuperCDMS SNOLAB to heavy DM but has no effect in the low-mass region that we are interested in. We divide the search window into 10 equally-spaced bins and include an energy resolution of $10\,\mathrm{keV}$. SuperCDMS expects to achieve a total exposure of $1.6 \cdot 10^4 \, \mathrm{kg\,days}$ over the course of five years of data taking. To compare CRESST-III and SuperCDMS on similar time-scales, however, we consider the sensitivity that can be achieved with a single year of data taking, corresponding to an exposure of approximately $3200 \, \mathrm{kg \, days}$.\footnote{Incidentally, this is also the exposure for which backgrounds become non-negligible (we expect about 2 events per bin), so increasing the exposure by a factor of 5 will improve the sensitivity by significantly less than a factor of 5.}

\subsection*{Xenon1T}

The first results of Xenon1T are based upon an exposure of $35636\, \mathrm{kg\,days}$~\cite{Aprile:2017iyp}. To calculate the acceptance function, we simulate fluctuations in the $S1$ and $S2$ signal making use of the scintillation and ionization yields from ref.~\cite{Akerib:2015rjg} and taking into account the anti-correlation between the two signals. Rather than attempting to model the detector response, we determine the light collection efficiency and the electron extraction probability by fitting to the nuclear recoil band shown in ref.~\cite{Aprile:2017iyp}. We then consider events with $S1 \geq 3\,\mathrm{phe}$ and require the $S2$ signal to lie below the mean of the nuclear recoil band. For low recoil energies, this procedure essentially reproduces the acceptance function shown in ref.~\cite{Aprile:2017iyp}, where no cut on the $S2$ signal is applied, whereas for large recoil energies we obtain an acceptance that is approximately a factor of 2 smaller. This result is in agreement with the expectation that the nuclear recoil cut has almost $100\%$ signal acceptance close to the threshold, where the signal results exclusively from upward fluctuations in the $S1$ signal, and around $50\%$ signal acceptance away from the threshold. The total number of expected background events below the mean of the nuclear recoil band is 0.36, and no events have been observed.
We therefore set a Poisson upper bound on the signal contribution of 1.94 events at 90\% CL, noting that this procedure leads to a slightly weaker bound than the asymptotic limit assumed in ref.~\cite{Aprile:2017iyp}.

\subsection*{LZ}

Our implementation of LZ is based upon the LZ conceptual design report~\cite{Akerib:2015cja}. Specifically, we assume a total exposure of $5.6 \cdot 10^6 \, \mathrm{kg\,days}$ and consider the search region $3\,\mathrm{phe} \leq S1 \leq 30\,\mathrm{phe}$. We calculate the energy-dependent acceptance by considering Poisson fluctuations in the $S1$ signal based on the effective scintillation yield for liquid xenon taken from the LUX analysis~\cite{Akerib:2015rjg} and assuming a light collection efficiency of $7.5\%$. We multiply this acceptance function by an overall factor of 1/2 to account for the requirement that the $S2$ signal must lie below the mean of the nuclear recoil band. Finally, we calculate the expected sensitivity by assuming that the number of observed events is equal to the expected background (2.37 events).

\section{Astrophysical uncertainties for low-mass dark matter}
\label{app:astro}

In the Galactic rest frame the standard Maxwell-Boltzmann velocity distribution is given by
\begin{equation}
 f(\mathbf{v}, v_0, v_\text{esc}) = \frac{1}{N} \left[ \exp\left(-\frac{v^2}{v_0^2}\right) - \exp\left(-\frac{v^2}{v_\text{esc}^2}\right) \right] \; ,
\end{equation}
where $N$ is an appropriate normalization constant. In the laboratory frame we need to account for the velocity of the Earth $\mathbf{v}_\mathrm{E}$ relative to the Galactic rest frame. Neglecting the motion of the Earth relative to the Sun, we take this velocity to be time-independent, $\mathbf{v}_\mathrm{E} = \mathbf{v}_\text{obs}$. We then obtain
\begin{align}
 f(v, \theta_v, v_0, v_\text{esc}, v_\text{obs}) = \frac{1}{N} \biggl[ & \exp\left(-\frac{v^2 + 2 \, v \, v_\text{obs} \cos \theta_v + v_\text{obs}^2}{v_0^2}\right) \nonumber \\
& - \exp\left(-\frac{v^2 + 2 \, v \, v_\text{obs} \cos \theta_v + v_\text{obs}^2}{v_\text{esc}^2}\right) \biggr] \; , \label{eq:SHM}
\end{align}
where $\theta_v$ denotes the angle between $\mathbf{v}_\text{obs}$ and the velocity vector $\mathbf{v}$.

Let us now consider two different velocity distributions of the Maxwell-Boltzmann form, $f(v)$ and $\tilde{f}(v)$, which are related by
\begin{equation}
 \tilde{v}_0 = v_0/z \, , \qquad \tilde{v}_\text{esc} = v_\text{esc}/z   \, , \qquad \tilde{v}_\text{obs} = v_\text{obs}/z \; ,
\end{equation}
where $z$ is an arbitrary rescaling factor.\footnote{We note that for an isothermal halo the velocity dispersion and the circular velocity are directly proportional to each other and it is therefore well-motivated to rescale $v_0$ and $v_\text{obs}$ simultaneously. The escape velocity is introduced by hand and is therefore in principle an independent parameter. Since our results do not depend sensitively on $v_\text{esc}$ we use the same rescaling factor for simplicity.} It then follows immediately from eq.~(\ref{eq:SHM}) that the two velocity distributions satisfy the relation
\begin{equation}
 \tilde{f}(v) = z^3 f(z v) \; ,
\end{equation}
where the pre-factor ensures that $\tilde{f}(v)$ is normalized, $\int f(v) \, \mathrm{d}^3 v  = \int \tilde{f}(v) \, \mathrm{d}^3 v = 1$. The velocity integral $\tilde{\eta}(v_\text{min})$ corresponding to $\tilde{f}(v)$ is then given by
\begin{align}
 \tilde{\eta}(v_\text{min}) & = \int_{v_\text{min}}^\infty \frac{\tilde{f}(v)}{v} \mathrm{d}^3 v = \int_{v_\text{min}}^\infty z^3 \frac{f(z v)}{v} \mathrm{d}^3 v = \int_{\tilde{v}_\text{min}}^\infty \alpha \frac{f(\tilde{v})}{\tilde{v}} \mathrm{d}^3 \tilde{v} = z \eta(\tilde{v}_\text{min}) \;,
\end{align}
where $\tilde{v}_\text{min} = z v_\text{min}$.

The differential event rate resulting from the rescaled velocity distribution therefore becomes
\begin{align}
\frac{\text{d}\tilde{R}_T}{\text{d}E_\mathrm{R}} = \frac{\rho_0 \, \xi_T}{2 \pi \, m_\text{DM}} \frac{g^2 \, F_T^2(E_\mathrm{R})}{\left( 2 \, m_T \, E_\mathrm{R} + m_\text{med}^2 \right)^2} \, z \eta (\tilde{v}_\text{min} (E_\mathrm{R}, m_\text{DM}))
\end{align}
The crucial observation is now that for $m_\text{DM} \ll m_T$ the minimum velocity is given by $v_\text{min}(E_\mathrm{R}, m_\text{DM}) = \frac{1}{m_\text{DM}} \sqrt{\frac{m_T \, E_\mathrm{R}}{2}}$
and hence $\tilde{v}_\text{min}(E_\mathrm{R}, m_\text{DM}) = {v}_\text{min}(E_\mathrm{R}, \tilde{m}_\text{DM})$ with $\tilde{m}_\text{DM} = m_\text{DM} / z$. In terms of this rescaled mass, the differential event rate becomes
\begin{align}
\frac{\text{d}\tilde{R}_T}{\text{d}E_\mathrm{R}} = \frac{\rho_0 \, \xi_T}{2 \pi \, \tilde{m}_\text{DM}} \frac{g^2 \, F_T^2(E_\mathrm{R})}{\left( 2 \, m_T \, E_\mathrm{R} + m_\text{med}^2 \right)^2} \, \eta (v_\text{min}(E_\mathrm{R}, \tilde{m}_\text{DM})) \; ,
\end{align}
which is identical to eq.~(\ref{eq:dRdE}) with $m_\text{DM}$ replaced by $\tilde{m}_\text{DM}$. In other words, as long as $m_\text{DM} \ll m_T$, the effect of an overall rescaling in the velocity distribution is equivalent to a rescaling in the DM mass.

\providecommand{\href}[2]{#2}\begingroup\raggedright\endgroup

\end{document}